\documentclass{elsart1p}

\usepackage{amsmath}
\usepackage{amsfonts}
\usepackage{amssymb}
\usepackage{array}
\usepackage{tikz}
\usetikzlibrary{shapes,arrows,automata}

{\newtheorem{theorem}{Theorem}}
{\newtheorem{corollary}{Corollary}}

\def\Nset{\mathbb{N}} 

\begin{document}

\begin{frontmatter}

\title{A boundary between universality and non-universality in spiking neural P systems\thanksref{ack}}

\author{Turlough Neary}
\address{Boole Centre for Research in Informatics, University  College Cork, Ireland.
}
\thanks[ack]{Turlough Neary is funded by Science Foundation Ireland Research Frontiers Programme grant number 07/RFP/CSMF641.}

\ead{tneary@cs.nuim.ie}
 \ead[url]{http://www.cs.nuim.ie/$\sim$tneary/}
\begin{abstract}
In this work we offer a significant improvement on the previous smallest spiking neural P systems and solve the problem of finding the smallest possible extended spiking neural P system. P\u{a}un and P\u{a}un~\cite{Paun2007} gave a universal spiking neural P system with 84 neurons and another that has extended rules with 49 neurons. Subsequently, Zhang et al.~\cite{Zhang2008B} reduced the number of neurons used to give universality to 67 for spiking neural P systems and to 41 for the extended model. Here we give a small universal spiking neural P system that has only 17 neurons and another that has extended rules with 5 neurons. All of the above mentioned spiking neural P systems suffer from an exponential slow down when simulating Turing machines. Using a more relaxed encoding technique we get a universal spiking neural P system that has extended rules with only 4 neurons. This latter spiking neural P system simulates 2-counter machines in linear time and thus suffer from a double exponential time overhead when simulating Turing machines. We show that extended spiking neural P systems with 3 neurons are simulated by log-space bounded Turing machines, and so there exists no such universal system with 3 neurons. It immediately follows that our 4-neuron system is the smallest possible extended spiking neural P system that is universal. Finally, we show that if we generalise the output technique we can give a universal spiking neural P system with extended rules that has only 3 neurons. This system is also the smallest of its kind as a universal spiking neural P system with extended rules and generalised output is not possible with 2 neurons.
\end{abstract}\vspace{-.4cm}

\end{frontmatter}

\section{Introduction}
Spiking neural P systems (SN P systems)~\cite{Ionescu2006} are quite a new computational model that are a synergy inspired by P systems and spiking neural networks. It has been shown that these systems are computationally universal~\cite{Ionescu2006}. Recently, P\u{a}un and P\u{a}un~\cite{Paun2007} gave two small universal SN P systems; They give an SN P system with 84 neurons and an extended SN P system with 49 neurons (that uses rules without delay). P\u{a}un and P\u{a}un conjectured that it is not possible to give a significant decrease in the number of neurons of their two universal systems. Zhang et al.~\cite{Zhang2008B} offered such a significant decrease in the number of neurons used to give such small universal systems. They give a universal SN P system with 67 neurons and another, which has extended rules (without delay), with 41 neurons. Here we give a small universal SN P system that has only 17 neurons and another, which has extended rules (without delay), with 5 neurons. Using a more relaxed encoding we get a universal SN P system that has extended rules (without delay), with 4 neurons. Table~\ref{tab:Small_SNP} gives the smallest universal SN P systems and their respective simulation time and space overheads. Note from Table~\ref{tab:Small_SNP} that, in addition to its small size, our 17-neuron system uses rules without delay. The other small universal SN P systems with standard rules~\cite{Paun2007,Zhang2008B} do not have this restriction.

In this work we also show that extended SN P systems with 3 neurons and generalised input are simulated by log-space bounded Turing machines. As a result, it is clear that there exists no such universal system with 3 neurons, and thus our 4-neuron system is the smallest possible universal extended SN P system. Following this, we show that if we generalise the output technique we can give a universal SN P system with extended rules that has only 3 neurons. In addition, we show that a universal SN P system with extended rules and generalised output is not possible with 2 neurons, and thus our 3-neuron systems is the smallest of its kind.

\begin{table}[h]
\begin{center}
\begin{tabular}{@{}c@{\:}|@{\:}c@{\:}|@{\;}c@{\;}|@{\;}c@{\,}|@{\,}c@{}}
number of & simulation & type  & exhaustive  & author \\
neurons & time/space &of rules& use of rules &\\ \hline
84	& exponential	& standard & no	& P{\u{a}}un and P{\u{a}}un~\cite{Paun2007}\\
67	& exponential	& standard & no & Zhang et al.~\cite{Zhang2008B}\\
49	& exponential	& extended\dag & no & P{\u{a}}un and P{\u{a}}un~\cite{Paun2007}\\
41	& exponential	& extended\dag & no & Zhang et al.~\cite{Zhang2008B}\\ 
12	& double-exponential	& extended\dag & no  & Neary~\cite{Neary2008B}\\
18	& exponential	& extended & no & Neary~\cite{Neary2008C,Neary2008B}*\\
125	& exponential/	& extended\dag & yes & Zhang et al.~\cite{Zhang2008A}\\
	& double-exponential	&  &  & \\
18	& polynomial/exponential	& extended & yes & Neary~\cite{Neary2008A}\\
10	& linear/exponential	& extended & yes &  Neary~\cite{Neary}\\ 
\textbf{17}	& \textbf{exponential}	& \textbf{standard\dag} & \textbf{no}	& \textbf{Section~\ref{sec:small SNP systems}}\\
\textbf{5}	& \textbf{exponential}	& \textbf{extended\dag} & \textbf{no}	& \textbf{Section~\ref{sec:small SNP systems}}\\
\textbf{4}	& \textbf{double-exponential}	& \textbf{extended\dag} & \textbf{no}	& \textbf{Section~\ref{sec:small SNP systems}}\\
\textbf{3}	& \textbf{double-exponential}	& \textbf{extended\ddag} & \textbf{no}	& \textbf{Section~\ref{sec:Lower bounds}}\\\hline

\end{tabular}
\end{center}
\caption{Small universal SN P systems. The ``simulation time'' column gives the overheads used by each system when simulating a standard single tape Turing machine. \dag~indicates that there is a restriction of the rules as delay is not used and \ddag~indicates that a more generalised output technique is used. *The 18 neuron system is not explicitly given in~\cite{Neary2008B}; it is however mentioned at the end of the paper and is easily derived from the other system presented in~\cite{Neary2008B}. Also, its operation and its graph were presented in~\cite{Neary2008C}.}\label{tab:Small_SNP}
\end{table}

From a previous result~\cite{Neary2008A} it is known that there exists no universal SN P system that simulates Turing machines in less the exponential time and space. It is a relatively straightforward matter to generalise this result to show that extended SN P systems suffer from the same inefficiencies. It immediately follows that the universal systems we present here and those found in~\cite{Paun2007,Zhang2008B} have exponential time and space requirements. However, it is possible to give a time efficient SN P system when we allow exhaustive use of rules. A universal extended SN P system with exhaustive use of rules has been given that simulates Turing machines in linear time~\cite{Neary}. Furthermore, this system has only 10 neurons. SN P systems with exhaustive use of rules were originally proved computationally universal by Ionescu et al.~\cite{Ionescu2007B}. However, the technique used to prove universality suffered from an exponential time overhead.

Using different forms of SN P systems, a number of time efficient (polynomial or constant time) solutions to NP-hard problems have been given~\cite{Chen2006A,Leporati2007A,Leporati2007B}. All of these solutions to NP-hard problems rely on families of SN P systems. Specifically, the size of the problem instance determines the number of neurons in the SN P system that solves that particular instance. This is similar to solving problems with circuits families where each input size has a specific circuit that solves it. Ionescu and Sburlan~\cite{Ionescu2007A} have shown that SN P systems simulate circuits in linear time. 

In Section~\ref{sec:small SNP systems} we give a definition for SN P systems, explain their operation and give other relevant technical details. In Section~\ref{sec:counter machines} we give a definition for counter machines and we also discuss some notions of universality. Following this, in Section~\ref{sec:small SNP systems} we give our small universal SN P systems and show how their size can be reduce if we use a more relaxed encoding. In Section~\ref{sec:Lower bounds} we give our proof showing that extended SN P systems with 3 neurons and generalised input are simulated by log-space bounded Turing machines. Section~\ref{sec:Lower bounds} also contains our universal 3-neuron system with generalised output. We end the paper with some discussion and conclusions. 

\section{SN P systems}\label{sec:Spiking neural P-systems}

\begin{defn}[Spiking neural P system]\label{def:Spiking neural P-systems}\hfill\\
A spiking neural P system (SN P system) is a tuple $\Pi=(O,\sigma_1,\sigma_2,\cdots,\sigma_m,syn,in,out)$, where:
	\begin{enumerate}
 		\item $O=\{s\}$ is the unary alphabet ($s$ is known as a spike),
 		\item $\sigma_1,\sigma_2,\cdots,\sigma_m$ are neurons, of the form $\sigma_i=(n_i,R_i),1\leqslant i\leqslant m$, where:
			\begin{enumerate}
 				\item $n_i\geqslant 0$ is the initial number of spikes contained in $\sigma_i$,
 				\item $R_i$ is a finite set of rules of the following two forms:
					\begin{enumerate}
 						\item $E/s^b\rightarrow s;d$, where $E$ is a regular expression over $s$, $b\geqslant 1$ and $d\geqslant 0$, 
 						\item $s^e\rightarrow\lambda$, where $\lambda$ is the empty word, $e\geqslant 1$, and for all $E/s^b\rightarrow s;d$ from $R_i$ $s^e\notin L(E)$ where $L(E)$ is the language defined by $E$, 
					\end{enumerate}
			\end{enumerate}
 		\item $syn\subseteq \{1,2,\cdots,m\}\times\{1,2,\cdots,m\}$ is the set of synapses between neurons, where $i\neq j$ for all  $(i,j)\in syn$, 
 		\item $in,out\in\{\sigma_1,\sigma_2,\cdots,\sigma_m\}$ are the input and output neurons, respectively.
	\end{enumerate}
\end{defn}\vspace{.3cm}
A firing rule $r=E/s^b\rightarrow s;d$ is applicable in a neuron $\sigma_i$ if there are $j\geqslant b$ spikes in $\sigma_i$ and $s^j\in L(E)$ where $L(E)$ is the set of words defined by the regular expression $E$. If, at time $t$, rule $r$ is executed then $b$ spikes are removed from the neuron, and at time $t+d$ the neuron fires. When a neuron $\sigma_i$ fires a spike is sent to each neuron $\sigma_j$ for every synapse $(i,j)$ in $\Pi$. Also, the neuron $\sigma_i$ remains closed and does not receive spikes until time $t+d$ and no other rule may execute in $\sigma_i$ until time $t+d+1$. A forgeting rule $r'=s^e\rightarrow\lambda$ is applicable in a neuron $\sigma_i$ if there are exactly $e$ spikes in $\sigma_i$. If $r'$ is executed then $e$ spikes are removed from the neuron. At each timestep $t$ a rule must be applied in each neuron if there is one or more applicable rules at time $t$. Thus, while the application of rules in each individual neuron is sequential the neurons operate in parallel with each other.

Note from 2b(i) of Definition~\ref{def:Spiking neural P-systems} that there may be two rules of the form $E/s^{b}\rightarrow s;d$, that are applicable in a single neuron at a given time. If this is the case then the next rule to execute is chosen non-deterministically.

An \emph{extended} SN P system~\cite{Paun2007} has more general rules of the form $E/s^b\rightarrow s^p;d$, where $b\geqslant p\geqslant 1$. Thus, a synapse in an SN P system with extended rules may transmit more than one spike in a single timestep. The SN P systems we present in this work use rules without delay, and thus in the sequel we write rules as $E/s^b\rightarrow s^p$. Also, if in a rule $E=s^b$ then we write the rule as $s^b\rightarrow s^p$.

In the same manner as in~\cite{Paun2007}, spikes are introduced into the system from the environment by reading in a binary sequence (or word) $w\in\{0,1\}$ via the input neuron $\sigma_1$. The sequence $w$ is read from left to right one symbol at each timestep and a spike enters the input neuron on a given timestep iff the read symbol is 1. The output of an SN P system $\Pi$ is the time between the first and second firing rule applied in the output neuron and is given by the value $\Pi(w)\in\Nset$.

A configuration $c$ of an SN P system consists of a word $w$ and a sequence of natural numbers $(r_1,r_2,\ldots ,r_m)$ where $r_i$ is the number of spikes in $\sigma_i$ and $w$ represents the remaining input yet to be read into the system. A computation step $c_j\vdash c_{j+1}$ is as follows: each number $r_i$ is updated depending on the number of spikes neuron $\sigma_i$ uses up and receives during the synchronous application of all applicable rules in configuration $c_j$. In addition, if $w\neq\lambda$ then the leftmost symbol of $w$ is removed. A SN P system computation is a finite sequence of configurations $c_1,c_2,\ldots,c_t$ that ends in a terminal configuration $c_t$ where for all $j<t$, $c_j\vdash c_{j+1}$. A \emph{terminal configuration} is a configuration where the input sequence has finished being read in via the input neuron (i.e. $w=\lambda$ the empty word) and either there is no applicable rule in any of the neurons or the output neuron has spiked exactly $v$ times (where $v$ is a constant independent of the input). 

Let $\phi_{x}$ be the $x^{th}$ $n$-ary partial recursive function in a G\"{o}del enumeration of all $n$-ary partial recursive functions. The natural number value $\phi_{x}(y_1,y_2,\ldots y_n)$ is the result given by $\phi_{x}$ on input $(y_1,y_2,\ldots y_n)$.

\begin{defn}\vspace{.3cm}[Universal SN P system]\label{def:Universal spiking neural P system}
A SN P system $\Pi$ is universal if there are recursive functions $g$ and $f$ such that for all $x,y\in\Nset$ we have $\phi_{x}(y_1,y_2,\ldots y_n)=f({\Pi}(g(x,y_1,y_2,\ldots y_n)))$.
\end{defn}\vspace{.3cm}
In the next section we give some further discussion on the subject of definitions of universality.

\section{Counter machines}\label{sec:counter machines} 
\begin{defn}[Counter machine]\label{def:counter machine}
A counter machine is a tuple $C=(z,R,c_{m},Q,q_1,q_h)$, where $z$ gives the number of counters, $R$ is the set of input counters, $c_{m}$ is the output counter, $Q=\{q_1,q_2,\cdots,q_h\}$ is the set of instructions, and $q_1,q_h\in Q$ are the initial and halt instructions, respectively.
\end{defn}\vspace{.3cm}
Each counter $c_j$ stores a natural number value $y\geqslant0$. Each instruction $q_i$ is of one of the following two forms $q_i:INC(j),q_{l}$ or $q_i:DEC(j),q_{l},q_k$ and is executed as follows:
\begin{itemize}
	\item $q_i:INC(j),q_{l}$ increment the value $y$ stored in counter $c_j$ by 1 and move to instruction $q_{l}$.
	\item $q_i:DEC(j),q_{l},q_k$ if the value $y$ stored in counter $c_j$ is greater than $0$ then decrement this value by 1 and move to instruction $q_{l}$, otherwise if $y=0$ move to instruction $q_{k}$.
\end{itemize}
At the beginning of a computation the first instruction executed is $q_1$. The input to the counter machine is initially stored in the input counters. If the counter machine's control enters instruction $q_h$, then the computation halts at that timestep. The result of the computation is the value $y$ stored in the output counter $c_m$ when the computation halts.

We now consider some different notions of universality. Korec~\cite{Korec1996} gives universality definitions that describe some counter machines as weakly universal and other counter machines as strongly universal. 

\begin{defn}\vspace{.3cm}[Korec~\cite{Korec1996}]\label{def:Korec strong universality}
A register machine $M$ will be called strongly universal if there is a recursive function $g$ such that for all $x,y\in\Nset$ we have $\phi_{x}(y)=\Phi_{M}^2(g(x),y)$.
\end{defn}\vspace{.3cm}
Here $\Phi_M^2(g(x),y)$ is the value stored in the output counter at the end of a computation when $M$ is started with the values $g(x)$ and $y$ in its input counters. Korec's definition insists that the value $y$ should not be changed before passing it as input to $M$. However, if we consider computing an $n$-arry function with a Korec-strong universal counter machine then it is clear that $n$ arguments must be encoded as a single input $y$. Many Korec-strong universal counter machines would not satisfy a definition where the function $\phi_{x}$ in Definition~\ref{def:Korec strong universality} is replaced with an $n$-arry function with $n>1$. For example, let us give a new definition where we replace the equation ``$\phi_{x}(y)=\Phi_M^2(g(x),y)$'' with the equation ``$\phi_{x}^n(y_1,y_2,\ldots,y_n)=\Phi_M^{n+1}(g(x),y_1,y_2,\ldots,y_n)$'' in Definition~\ref{def:Korec strong universality}. Note that for any counter machine $M$ with $r$ counters, if $r\leqslant n$ then $M$ does not satisfy this new definition. It could be considered that Korec's notion of strong universality is somewhat arbitrary for the following reason: Korec's definition will admit machines that require $n$-arry input $(y_1,y_2,\ldots,y_n)$ to be encoded as the single input $y$ when simulating an $n$-arry function, but his definition will not admit a machine that applies an encoding function to $y$ (e.g. $y^2$ is not permitted). Perhaps when one uses this notion of universality it would be more appropriate to refer to it as strongly universal for unary partial recursive functions instead of simply strongly universal.

Korec~\cite{Korec1996} also gives a number of other definitions of universality. If the equation $\phi_{x}(y)=\Phi_M^2(g(x),y)$ in Definition~\ref{def:Korec strong universality} above is replaced with any one of the equations $\phi_{x}(y)=\Phi_M^1(g_2(x,y))$, $\phi_{x}(y)=f(\Phi_M^2(g(x),y))$ or $\phi_{x}(y)=f(\Phi_M^1(g_2(x,y)))$ then the counter machine $M$ is weakly universal. Korec gives another definition where the equation $\phi_{x}(y)=\Phi_M^2(g(x),y)$ in Definition~\ref{def:Korec strong universality} is replaced with the equation $\phi_{x}(y)=f(\Phi_M^2(g(x),h(y)))$. However, he does not include this definition in his list of weakly universal machines even though the equation $\phi_{x}(y)=f(\Phi_M^2(g(x),h(y)))$ allows for a more relaxed encoding than the
equation $\phi_{x}(y)=f(\Phi_M^2(g(x),y))$ and thus gives a weaker form of universality. 

For each number $m>2$ there exists universal $m$-counter machines that allow $\phi_x^n$ and its input $(y_1,y_2,\ldots,y_n)$ to be encoded separately (e.g. via $g(x)$ and $h^n(y_1,y_2,\ldots,y_n)$). For universal 2-counter machines all of the current algorithms encode the function $\phi_x^n$ and its input $(y_1,y_2,\ldots,y_n)$ together as a single input (e.g. via $g^{n+1}(x,y_1,y_2,\ldots,y_n)$). Using such encodings it is only possible to give universal 2-counter machines that Korec would class as weakly universal. Some other limitations of 2-counter machines were shown independently by Schroeppel~\cite{Schroeppel1972} and Barzdin~\cite{Barzdin1963}. In both cases the authors are examining unary functions that are uncomputable for 2-counter machines when the input value to the counter machine must equal the input to the function. For example Schroeppel shows that given $n$ as input a 2-counter machine cannot compute $2^n$. It is interesting to note that one can give a Korec-strong universal counter machine that is as time/space inefficient as a Korec-weak universal 2-counter machine. Korec's definition of strong universality deals with input and output only and is not concerned with the (time/space) efficiency of the computation.

In earlier work~\cite{Paun2007}, Korec's notion of strong universality was adopted for SN P systems\footnote{Note that no formal definition of this notion was explicitly given in\cite{Paun2007}.} as follows: A spiking neural P system $\Pi$ is strongly universal if $\Pi(10^{y-1}10^{x-1}1)=\phi_x(y)$ for all $x$ and $y$ (here if $\phi_x(y)$ is undefined so to is $\Pi(10^{y-1}10^{x-1}1)$). As with the SN P systems given in~\cite{Paun2007,Zhang2008B}, the systems we give in Theorems~\ref{thm:universal extended SNP systems with 5 neurons} and~\ref{thm:universal SNP systems with 17 neuron} satisfy the notion of strong universality adopted from Korec in~\cite{Paun2007}. Analogously, our system in Theorem~\ref{thm:universal Extended SNP system with 4 neuron} could be compared to what Korec refers to as weak universality. However, as we noted in our analysis above, it could be considered that Korec's notion of strong universality is somewhat arbitrary and we also pointed out some inconsistency in his notion of weak universality. Hence, in this work we rely on time/space complexity analysis to compare the encodings used by small SN P system (see Table~\ref{tab:Small_SNP}).

It is well known that counter machines require an exponential time overhead to simulate Turing machines~\cite{Fischer1968}. Counter machines with only 2 counters are universal~\cite{Minsky1967}, however, they simulate Turing machines with a double exponential time overhead. In the sequel we give some universal SN P systems that simulate 3-counter machines and others that simulate 2-counter machines. The reason for this is that when using our algorithm there is a trade-off between the size and the time efficiency of the system. This trade-off is dependant on whither we choose to simulate 3-counter machines or 2-counter machines. When simulating Turing machines, 3-counter machines suffer from an exponential time overhead and 2-counter machines suffer from a double-exponential time overhead, and thus the simulation of 3-counter machines is preferable when considering the time efficiency of the system. If we are considering the size of our system then 2-counter machines have an advantage over 3-counter machines as our algorithms require a constant number of neurons to simulate each counter. 

\section{Small universal SN P systems}\label{sec:small SNP systems}
We begin this section by giving our two extended universal systems $\Pi_{C_3}$ and $\Pi_{C_2}$, and following this we give our standard system $\Pi'_{C_3}$. We prove the universality of $\Pi_{C_3}$ and $\Pi'_{C_3}$ by showing that they each simulate a universal 3-counter machine. From $\Pi_{C_3}$ we obtain the system $\Pi'_{C_2}$ which simulates a universal 2-counter machine.

\begin{theorem}\label{thm:universal extended SNP systems with 5 neurons}
Let $C_3$ be a universal counter machine with 3 counters that completes it computation in time $t$ to give the output value $x_o$ when given the pair of input values ($x_1$, $x_2$). Then there is a universal extended SN P system $\Pi_{C_3}$ that simulates the computation of $C_3$ in time $O(t+x_1+x_2+x_o)$ and has only 5 neurons. 
\end{theorem}

\begin{figure}[t]
\begin{center}
\begin{tikzpicture}[>=stealth',shorten >=1pt,auto,node distance=1.5cm,thick,bend angle=45]

	\tikzstyle{dots}=[draw=none,node distance=1cm]
	\tikzstyle{state}=[ellipse,draw=black!75, node distance=1.2cm]
	\tikzstyle{state1}=[ellipse,draw=black!75, node distance=2.4cm]
	\tikzstyle{state2}=[ellipse,draw=black!75, node distance=1.6cm]
	\tikzstyle{state3}=[ellipse,draw=black!75, node distance=.2cm]
	\tikzstyle{input-output}=[draw=none,node distance=1cm]
	\tikzstyle{dummy1}=[draw=none,node distance=.2cm]

	\node[state1](sigma3)[]	{\footnotesize counter $c_2$};	\draw (sigma3)+(-.7,-.44) node {$\sigma_3$};
	\node[state](sigma5)[below of=sigma3]	{$\quad^{ }$};	\draw (sigma5)+(0,0) node {$\sigma_{5}$};
	\node[state1](sigma2)[left of=sigma3]	{\footnotesize counter $c_1$};	\draw (sigma2)+(-1.4,0) node {$\sigma_2$};
	\node[state1](sigma4)[right of=sigma3]	{\footnotesize counter $c_3$};	\draw (sigma4)+(1.4,0) node {$\sigma_4$};
	\node[dummy1](dummy1)[above of=sigma3]		{};	
	\node[state2](sigma1)[above of=dummy1]		{$\quad^{ }$};			\draw (sigma1)+(0,0) node {$\sigma_{1}$};
	\node[input-output](input)[above of=sigma1]	{};
	\node[input-output](output)[below of=sigma5]	{};

\path[->]	(input)edge	node 	{\small input} (sigma1)
		(sigma1)edge	node 	{}      (sigma2)
		(sigma1)edge	node 	{}      (sigma3)
		(sigma1)edge	node 	{}      (sigma4)
		(sigma2)edge	node	{}	(sigma1)
		(sigma3)edge	node	{}	(sigma1)
		(sigma3)edge	node	{}	(sigma5)
		(sigma4)edge	node	{}	(sigma1)
		(sigma5)edge	node {\small output} (output);

	\end{tikzpicture}
            \end{center}
            \caption{Universal extended SN P system $\Pi_{C_3}$. Each oval labeled $\sigma_i$ is a neuron. An arrow going from neuron $\sigma_i$ to neuron $\sigma_j$ illustrates a synapse $(i,j)$.}\label{fig:extended universal SNP system}
        \end{figure}
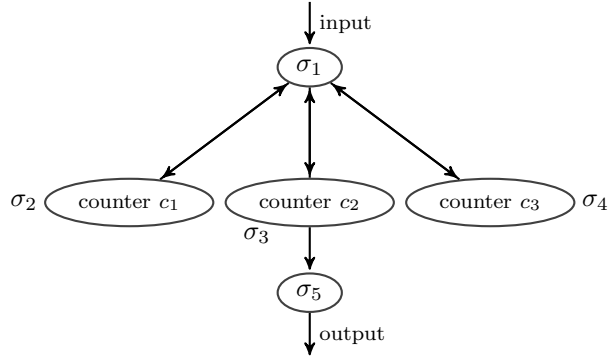

\begin{pf}
Let $C_3=(3,\{c_{1},c_{2}\},c_2,Q,q_1,q_h)$ where $Q=\{q_1,q_2,\cdots,q_h\}$. Our SN P system $\Pi_{C_3}$ is given by Figure~\ref{fig:extended universal SNP system} and Table~\ref{tab:neurons of Extended SNP 3-counter}. The algorithm given for $\Pi_{C_3}$ is deterministic. 

\subsubsection{Encoding of a configuration of $C_3$ and reading input into $\Pi_{C_3}$} A configuration of $C_3$ is stored as spikes in the neurons of $\Pi_{C_3}$. The next instruction $q_i$ to be executed is stored in each of the neurons $\sigma_2$, $\sigma_{3}$ and $\sigma_{4}$ as $4(h+i)$ spikes. Let $x_1$, $x_2$ and $x_3$ be the values stored in counters $c_1$, $c_2$ and $c_3$, respectively. Then the values $x_1$, $x_2$ and $x_3$ are stored as $8h(x_1+1)$, $8h(x_2+1)$ and $8h(x_3+1)$ spikes in neurons $\sigma_2$, $\sigma_{3}$ and $\sigma_{4}$, respectively.

The input to $\Pi_{C_3}$ is read into the system via the input neuron $\sigma_1$ (see Figure~\ref{fig:extended universal SNP system}). If $C_3$ begins its computation with the values $x_1$ and $x_2$ in counters $c_1$ and $c_2$, respectively, then the binary sequence $w=10^{x_1-1}10^{x_2-1}1$ is read in via the input neuron $\sigma_1$. Thus, $\sigma_1$ receives a single spike from the environment at times $t_1$, $t_{x_1+1}$ and $t_{x_1+x_2+1}$. We explain how the system is initialised to encode an initial configuration of $C_3$ by giving the number of spikes in each neuron and the rule that is to be applied in each neuron at time $t$. Before the computation begins neuron $\sigma_1$ initially contain $8h$ spikes, $\sigma_3$ contains $2$ spikes, $\sigma_4$ contains $8h+1$ spikes and all other neurons contain no spikes. Thus, when $\sigma_1$ receives it first spike at time $t_1$ we have
\begin{xalignat*}{2}
&t_{1}:\; \\
	&\qquad\sigma_1=8h+1,	& s^{8h+1}/s^{8h}\rightarrow s^{8h}, \\
	&\qquad\sigma_3=2,	& s^2/s\rightarrow s, \\
	&\qquad\sigma_4=8h+1,	& s^{8h+1}/s^{8h}\rightarrow s^{8h-1}.
\end{xalignat*}
where on the left $\sigma_k=z$ gives the number $z$ of spikes in neuron $\sigma_k$ at time $t$ and on the right is the rule that is to be applied at time $t$, if there is an applicable rule at that time. Thus, from Figure~\ref{fig:extended universal SNP system}, when we apply the rule $s^{8h+1}/s^{8h}\rightarrow s^{8h}$ in neuron $\sigma_1$, $s^2/s\rightarrow s$ in $\sigma_3$, and $s^{8h+1}/s^{8h}\rightarrow s^{8h-1}$ in $\sigma_4$ at time $t_1$ we get
\begin{xalignat*}{2}
&t_{2}:\; \\
	&\qquad\sigma_1=8h+1,	& s^{8h+1}/s^{8h}\rightarrow s^{8h}, \\
	&\qquad\sigma_2=8h,	& \\
	&\qquad\sigma_3=8h+1,	& s^{8h+1}/s^{8h}\rightarrow s, \\
	&\qquad\sigma_4=8h+1,	& s^{8h+1}/s^{8h}\rightarrow s^{8h-1},\\
	&\qquad\sigma_5=1,	& s\rightarrow \lambda,\\
\\
&t_{3}:\; \\
	&\qquad\sigma_1=8h+1,	& s^{8h+1}/s^{8h}\rightarrow s^{8h}, \\
	&\qquad\sigma_2=16h,	& \\
	&\qquad\sigma_3=8h+1,	& s^{8h+1}/s^{8h}\rightarrow s, \\
	&\qquad\sigma_4=8h+1,	& s^{8h+1}/s^{8h}\rightarrow s^{8h-1},\\
	&\qquad\sigma_5=1,	& s\rightarrow \lambda.
\end{xalignat*}
Neuron $\sigma_1$ fires on every timestep between times $t_1$ and $t_{x_1+1}$ to send a total of $8hx_1$ spikes to $\sigma_2$ thus we get
\begin{xalignat*}{2}
&t_{x_1+1}:\; \\
	&\qquad\sigma_1=8h+2,	& s^{8h+2}/s^{8h+1}\rightarrow s^{8h+1}, \\
	&\qquad\sigma_2=8hx_1,	& \\
	&\qquad\sigma_3=8h+1,	& s^{8h+1}/s^{8h}\rightarrow s, \\
	&\qquad\sigma_4=8h+1,	& s^{8h+1}/s^{8h}\rightarrow s^{8h-1},\\
	&\qquad\sigma_5=1,	& s\rightarrow \lambda,\\
\\
&t_{x_1+2}:\; \\
	&\qquad\sigma_1=8h+1,	& s^{8h+1}/s^{8h}\rightarrow s^{8h}, \\
	&\qquad\sigma_2=8h(x_1+1)+1,	& (s^{8h})^{\ast}s^{8h+1}/s^{8h}\rightarrow s,\\
	&\qquad\sigma_3=8h+2,	&  \\
	&\qquad\sigma_4=8h+2,	& s^{8h+2}/s^{8h}\rightarrow s^{8h-1},\\
	&\qquad\sigma_5=1,	& s\rightarrow \lambda,\\
\\
&t_{x_1+3}:\; \\
	&\qquad\sigma_1=8h+1,	& s^{8h+1}/s^{8h}\rightarrow s^{8h}, \\
	&\qquad\sigma_2=8h(x_1+1)+1,	& (s^{8h})^{\ast}s^{8h+1}/s^{8h}\rightarrow s,\\
	&\qquad\sigma_3=16h+2,	&  \\
	&\qquad\sigma_4=8h+2,	& s^{8h+2}/s^{8h}\rightarrow s^{8h-1}.
\end{xalignat*}
Neuron $\sigma_1$ fires on every timestep between times $t_{x_1+1}$ and $t_{x_1+x_2+1}$ to send a total of $8hx_2$ spikes to $\sigma_3$. Thus, when $\sigma_1$ receives the last spike from its environment we have
\begin{xalignat*}{2}
&t_{x_1+x_2+1}:\; \\
	&\qquad\sigma_1=8h+2,	& s^{8h+2}/s^{8h+1}\rightarrow s^{8h+1}, \\
	&\qquad\sigma_2=8h(x_1+1)+1,	& (s^{8h})^{\ast}s^{8h+1}/s^{8h}\rightarrow s,\\
	&\qquad\sigma_3=8hx_2+2,	&  \\
	&\qquad\sigma_4=8h+2,	& s^{8h+2}/s^{8h}\rightarrow s^{8h-1}\\
\\
&t_{x_1+x_2+2}:\; \\
	&\qquad\sigma_1=8h+1,	& s^{8h+1}/s^{8h}\rightarrow s^{8h}, \\
	&\qquad\sigma_2=8h(x_1+1)+2,	& (s^{8h})^{\ast}s^{8h+2}/s^{8h+2}\rightarrow s^{2h},\\
	&\qquad\sigma_3=8h(x_2+1)+3,		& (s^{8h})^{\ast}s^{8h+3}/s^{8h+3}\rightarrow s^{2h}, \\
	&\qquad\sigma_4=8h+3,	& s^{8h+3}\rightarrow s^{2h}.
\end{xalignat*}
\begin{xalignat*}{2}
&t_{x_1+x_2+3}:\; \\
	&\qquad\sigma_1=6h+1,		& s^{6h+1}\rightarrow s^{4h+4}, \\
	&\qquad\sigma_2=8h(x_1+1),	& \\
	&\qquad\sigma_3=8h(x_2+1),		& \\
	&\qquad\sigma_4=8h,	& \\
	&\qquad\sigma_5=2h,		& s^{2h}\rightarrow \lambda,\\
\\
&t_{x_1+x_2+4}:\; \\
	&\qquad\sigma_2=8h(x_1+1)+4(h+1),	& \\
	&\qquad\sigma_3=8h(x_2+1)+4(h+1),		& \\
	&\qquad\sigma_4=8h+4(h+1).	&
\end{xalignat*}
At time $t_{x_1+x_2+4}$ neuron $\sigma_2$ contains $8h(x_1+1)+4(h+1)$ spikes, $\sigma_3$ contains $8h(x_2+1)+4(h+1)$ spikes and $\sigma_4$ contains $8h+4(h+1)$ spikes. Thus at time $t_{x_1+x_2+4}$ the SN P system encodes an initial configuration of $C_3$.

\subsubsection{$\Pi_{C_3}$ simulating $q_i:INC(1),q_{l}$} 
Let counters $c_1$, $c_2$, and $c_3$ have values $x_1$, $x_2$, and $x_3$, respectively. Then the simulation of $q_i:INC(1),q_{l}$ begins at time $t_j$ with $8h(x_1+1)+4(h+i)$ spikes in $\sigma_{2}$, $8h(x_2+1)+4(h+i)$ spikes in $\sigma_{3}$ and $8h(x_3+1)+4(h+i)$ spikes in $\sigma_{4}$. Thus, at time $t_j$ we have
\begin{xalignat*}{2}
&t_{j}:\; \\
	&\qquad\sigma_2=8h(x_1+1)+4(h+i),	& (s^{8h})^{\ast}s^{4(h+i)}/s^{4(h+i)}\rightarrow s^{4(h+i)}, \\
        &\qquad\sigma_3=8h(x_2+1)+4(h+i),	& (s^{8h})^{\ast}s^{4(h+i)}/s^{8h+4(h+i)}\rightarrow s^{6h}, \\
        &\qquad\sigma_4=8h(x_3+1)+4(h+i),	& (s^{8h})^{\ast}s^{4(h+i)}/s^{8h+4(h+i)}\rightarrow s^{6h}. 
\end{xalignat*}
From Figure~\ref{fig:extended universal SNP system}, when we apply the rule $(s^{8h})^{\ast}s^{4(h+i)}/s^{4(h+i)}\rightarrow s^{4(h+i)}$ in neuron $\sigma_2$ and the rule $(s^{8h})^{\ast}s^{4(h+i)}/s^{8h+4(h+i)}\rightarrow s^{6h}$ in $\sigma_{3}$ and $\sigma_{4}$ at time $t_j$ we get
\begin{xalignat*}{2}
&t_{j+1}:\; \\
	&\qquad\sigma_{1}=16h+4i,	& s^{16h+4i}\rightarrow s^{12h+4l}, \\
	&\qquad\sigma_2=8h(x_1+1),	& \\
        &\qquad\sigma_3=8hx_2,		& \\
        &\qquad\sigma_4=8hx_3,		& \\ 
        &\qquad\sigma_{5}=6h,		& s^{6h}\rightarrow \lambda, \\
\\
&t_{j+2}:\; \\
	&\qquad\sigma_2=8h(x_1+2)+4(h+l),	& \\
	&\qquad\sigma_3=8h(x_2+1)+4(h+l),		& \\
	&\qquad\sigma_4=8h(x_3+1)+4(h+l),		& \\ 
\end{xalignat*}
At time $t_{j+2}$ the simulation of $q_i:INC(1),q_{l}$ is complete. Note that an increment on the value $x_1$ in counter $c_1$ was simulated by increasing the $8h(x_1+1)$ spikes in $\sigma_{2}$ to $8h(x_1+2)$ spikes. Note also that the encoding $4(h+l)$ of the next instruction $q_{l}$ has been established in neurons $\sigma_{2}$, $\sigma_{3}$ and $\sigma_{4}$.

\subsubsection{$\Pi_{C_3}$ simulating $q_i:DEC(1),q_{l},q_k$} 
There are two cases to consider here. Case 1: if counter $c_1$ has value $x_1>0$, then decrement counter 1 and move to instruction $q_{i+1}$. Case 2: if counter $c_1$ has value $x_1=0$, then move to instruction $q_{k}$. As with the previous example, our simulation begins at time $t_j$. Thus Case 1 ($x_1>0$) gives
\begin{xalignat*}{2}
&t_{j}:\; \\
	&\qquad\sigma_2=8h(x_1+1)+4(h+i),	& (s^{8h})^{\ast}s^{16h+4(h+i)}/s^{12h+4i}\rightarrow s^{6h+4i}, \\
        &\qquad\sigma_3=8h(x_2+1)+4(h+i),	& (s^{8h})^{\ast}s^{4(h+i)}/s^{4(h+i)}\rightarrow s^{2h}, \\
        &\qquad\sigma_4=8h(x_3+1)+4(h+i),	& (s^{8h})^{\ast}s^{4(h+i)}/s^{4(h+i)}\rightarrow s^{2h}, \\
\\
&t_{j+1}:\; \\
	&\qquad\sigma_{1}=10h+4i,	& s^{10h+4i}\rightarrow s^{4(h+l)}, \\
	&\qquad\sigma_2=8hx_1,	& \\
        &\qquad\sigma_3=8h(x_2+1),	& \\
        &\qquad\sigma_4=8h(x_3+1),	& \\ 
	&\qquad\sigma_{5}=2h,	& s^{2h}\rightarrow \lambda, \\
\\
&t_{j+2}:\; \\
	&\qquad\sigma_2=8hx_1+4(h+l),	& \\
        &\qquad\sigma_3=8h(x_2+1)+4(h+l),	& \\
        &\qquad\sigma_4=8h(x_3+1)+4(h+l).	& 
\end{xalignat*}
At time $t_{j+2}$ the simulation of $q_i:DEC(1),q_{l},q_k$ for Case 1 ($x_1>0$) is complete. Note that a decrement on the value $x_1$ in counter $c_1$ was simulated by decreasing the $8h(x_1+1)$ spikes in $\sigma_{2}$ to $8hx_1$ spikes. Note also that the encoding $4(h+l)$ of the next instruction $q_{l}$ has been established in neurons $\sigma_{2}$, $\sigma_{3}$ and $\sigma_{4}$. Alternatively, if we have Case 2 ($x_1=0$) then we get 
\begin{xalignat*}{2}
&t_{j}:\; \\
	&\qquad\sigma_2=8h+4(h+i),		& s^{8h+4(h+i)}/s^{4(h+i)}\rightarrow s^{4(h+i)}, \\
        &\qquad\sigma_3=8h(x_2+1)+4(h+i),	& (s^{8h})^{\ast}s^{4(h+i)}/s^{4(h+i)}\rightarrow s^{2h}, \\
        &\qquad\sigma_4=8h(x_3+1)+4(h+i),	& (s^{8h})^{\ast}s^{4(h+i)}/s^{4(h+i)}\rightarrow s^{2h}, \\
\\
&t_{j+1}:\; \\
        &\qquad\sigma_{1}=8h+4i,	& s^{8h+4i}\rightarrow s^{4(h+k)}, \\
	&\qquad\sigma_2=8h,		&\\
        &\qquad\sigma_3=8h(x_2+1),	& \\
        &\qquad\sigma_4=8h(x_3+1),	& \\ 
	&\qquad\sigma_{5}=2h,		& s^{2h}\rightarrow \lambda.
\end{xalignat*}
\begin{xalignat*}{2}
&t_{j+2}:\; \\
	&\qquad\sigma_2=8h+4(h+k),	& \\
        &\qquad\sigma_3=8h(x_2+1)+4(h+k),	& \\
        &\qquad\sigma_4=8h(x_3+1)+4(h+k).	& 
\end{xalignat*}
At time $t_{j+2}$ the simulation of $q_i:DEC(1),q_l,q_k$ for Case 1 ($x_1=0$) is complete. The encoding $4(h+k)$ of the next instruction $q_{k}$ has been established in neurons $\sigma_{2}$, $\sigma_{3}$ and $\sigma_{4}$.

\subsubsection{Halting}
The halt instruction $q_h$ is encoded as $4h+5$ spikes. Thus, if $C_3$ enters the halt instruction $q_h$ we get
\begin{xalignat*}{2}
&t_{j}:\; \\
	&\qquad\sigma_2=8h(x_1+1)+4h+5,		& \\
        &\qquad\sigma_3=8h(x_o+1)+4h+5,		& (s^{8h})^{\ast}s^{20h+5}/s^{12h}\rightarrow s^2, \\
        &\qquad\sigma_4=8h(x_3+1)+4h+5,		& \\
\\
&t_{j+1}:\; \\
        &\qquad\sigma_{1}=2,			& s^{2}\rightarrow \lambda,\\
	&\qquad\sigma_2=8h(x_1+1)+4h+5,		& \\
        &\qquad\sigma_3=8hx_o+5,		& (s^{8h})^{\ast}s^{16h+5}/s^{8h}\rightarrow s, \\
        &\qquad\sigma_4=8h(x_3+1)+4h+5,		& \\
        &\qquad\sigma_{5}=2,			& s^{2}\rightarrow s,\\
\\
&t_{j+2}:\; \\
        &\qquad\sigma_{1}=1,			& s\rightarrow \lambda,\\
	&\qquad\sigma_2=8h(x_1+1)+4h+5,		& \\
        &\qquad\sigma_3=8h(x_o-1)+5,		& (s^{8h})^{\ast}s^{16h+5}/s^{8h}\rightarrow s, \\
        &\qquad\sigma_4=8h(x_3+1)+4h+5,		& \\
        &\qquad\sigma_{5}=1,			& s\rightarrow \lambda.
\end{xalignat*}
The rule $(s^{8h})^{\ast}s^{16h+5}/s^{8h}\rightarrow s$ is applied a further $x_o-2$ times in $\sigma_{3}$ until we get
\begin{xalignat*}{2}
&t_{j+x_o}:\; \\
        &\qquad\sigma_{1}=1,			& s\rightarrow \lambda,\\
	&\qquad\sigma_2=8h(x_1+1)+4h+5,		& \\
        &\qquad\sigma_3=8h+5,			& s^{8h+5}\rightarrow s^2, \\
        &\qquad\sigma_4=8h(x_3+1)+4h+5,		& \\
        &\qquad\sigma_{5}=1,			& s\rightarrow \lambda.
\end{xalignat*}
\begin{xalignat*}{2}
&t_{j+x_o+1}:\; \\
        &\qquad\sigma_{1}=2,			& s^{2}\rightarrow \lambda,\\
	&\qquad\sigma_2=8h(x_1+1)+4h+5,		& \\
        &\qquad\sigma_4=8h(x_3+1)+4h+5,		& \\
        &\qquad\sigma_{5}=2,			& s^2\rightarrow s.\\
\end{xalignat*}
As usual the output is the time interval between the first and second spikes that are sent out of the output neuron. Note from above that the output neuron $\sigma_{5}$ fires for the first time at timestep $t_{j+1}$ and for the second time at timestep $t_{j+x_o+1}$. Thus, the output of $\Pi_{C_3}$ is $x_o$ the value of the output counter $c_2$ when $C_3$ enters the halt instruction $q_h$. Note that if $x_2=0$ then the rule $s^{12h+5}\rightarrow s^2$ is executed at timestep $t_{j}$, and thus only one spike will be sent out of the output neuron.

We have now shown how to simulate arbitrary instructions of the form $q_i:INC(1),q_{l}$ and ${q_i:DEC(1),q_{l},q_k}$ that operate on counter $c_1$. Instructions which operate on counters $c_2$ and $c_3$ are simulated in a similar manner. Immediately following the simulation of an instruction $\Pi_{C_3}$ is configured to simulate the next instruction. Each instruction of $C_3$ is simulated in 2 timesteps. The pair of input values ($x_1,x_2$) is read into the system in $x_1+x_2+4$ timesteps and sending the output value $x_o$ out of the system takes $x_o+1$ timesteps. Thus, if $C_3$ completes it computation in time $t$, then $\Pi_{C_3}$ simulates the computation of $C_3$ in linear time $O(t+x_1+x_2+x_o)$. \qed
\end{pf}

\begin{theorem}\label{thm:universal Extended SNP system with 4 neuron}
Let $C_2$ be a universal counter machine with 2 counters that completes it computation in time $t$ to give the output value $x_o$ when given the input value $x_1$. Then there is a universal extended SN P system $\Pi_{C_2}$ that simulates the computation of $C_2$ in time $O(t+x_1+x_o)$ and has only 4 neurons. 
\end{theorem}
\begin{pf}
Let $C_2=(2,\{c_{1}\},c_{2},Q,q_1,q_h)$ where $Q=\{q_1,q_2,\cdots,q_h\}$. The rules for the SN P system $\Pi_{C_2}$ are given by Table~\ref{tab:neurons of Extended SNP 2-counter} and a diagram of the system is obtained by removing neuron $\sigma_4$ from Figure~\ref{fig:extended universal SNP system}. If $C_2$ begins its computation with the value $x_1$ in counter $c_1$ then the binary sequence $w=10^{x_1-1}1$ is read in via the input neuron $\sigma_1$. Before the computation begins neurons $\sigma_1$, $\sigma_2$, $\sigma_3$ and $\sigma_5$ respectively contain $8h$, $8h+1$, $16h+1$ and 0 spikes. Like $\Pi_{C_3}$, $\Pi_{C_2}$ encodes the value $x$ of each counter as $8h(x+1)$ spikes and encodes each instruction $q_i$ as $4(h+i)$ spikes. The operation of $\Pi_{C_2}$ is very similar to the operation of $\Pi_{C_3}$, and thus it would be tedious and repetitive to go through another simulation here. $\Pi_{C_2}$ simulates a single instruction of $C_2$ in $2$ timesteps in a manner similar to that of $\Pi_{C_3}$. The inputting and outputting techniques, used by $\Pi_{C_2}$, also remain similar to those of $\Pi_{C_3}$, and thus the running time of $\Pi_{C_2}$ is $O(t+x_1+x_o)$. \qed
\end{pf}

The SN P system in Theorem~\ref{thm:universal SNP systems with 17 neuron} simulates a counter machine with the following restriction: if a counter is being decremented no other counter has value 0 at that timestep. Note that this does not result in a loss of generality as for each standard counter machine there is a counter machine with this restriction that simulates it in linear time without an increase in the number of counters. Let $C$ be any counter machine with $m$ counters. Then there is a counter machine $C'$ with $m$ counters that simulates $C$ in linear time, such that if $C'$ is decrementing a counter no other counter has value 0 at that timestep. Each counter in $C$ that has value $y$ is simulated by a counter in $C'$ that has value $y+1$. The instruction set of $C'$ is the same as the instruction set of $C$ with the following exception each $q_i:DEC(j),q_l,q_k$ instruction in $C$ is replaced with the instructions $(q_i:DEC(j)q'_{i},q'_{i})$, $(q'_{i}:DEC(j)q^{\star}_l,q^{\star}_{k})$, $(q^\star_{l}:INC(j),q_l)$, and $(q^{\star}_{k}:INC(j),q_k)$. The reason we need these extra instructions is that $y$ is encoded as $y+1$ and we must decrement twice if we wish to test for an encoded 0.

\begin{theorem}\vspace{.3cm}\label{thm:universal SNP systems with 17 neuron}
Let $C_3$ be a universal counter machine with 3 counters and $h$ instructions that completes it computation in time $t$ to give the output value $x_o$ when given the input $(x_1,x_2)$. Then there is a universal SN P system $\Pi'_{C_3}$ that simulates the computation of $C_3$ in time $O(ht+x_1+x_2+x_o)$ and has only 17 neurons. 
\end{theorem}

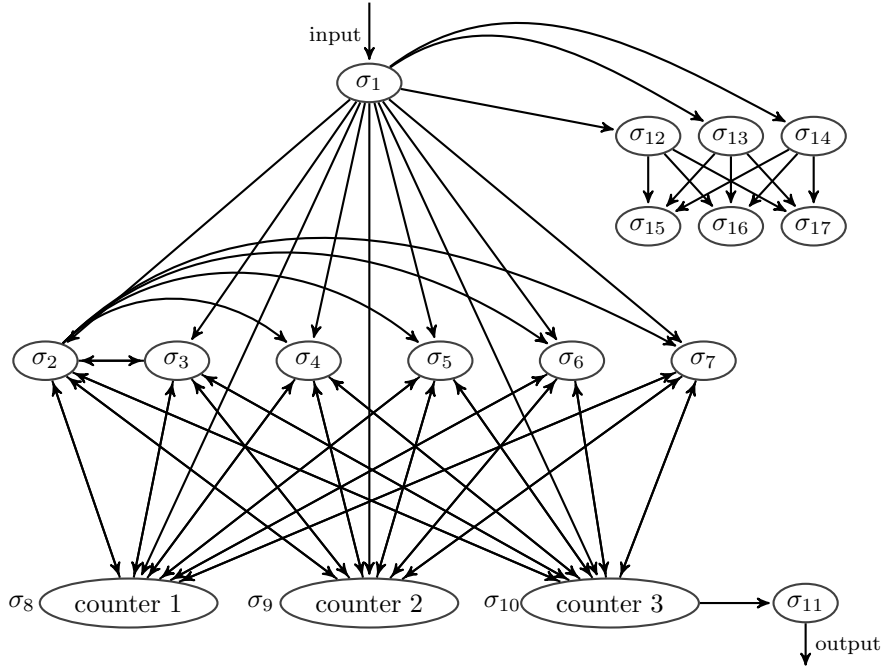
\begin{figure}[t]
\begin{center}
\begin{tikzpicture}[>=stealth',shorten >=1pt,auto,node distance=1.5cm,thick,bend angle=45]

	\tikzstyle{dots}=[draw=none,node distance=1cm]
	\tikzstyle{state}=[ellipse,draw=black!75, node distance=1.4cm]
	\tikzstyle{state1}=[ellipse,draw=black!75, node distance=1.1cm]
	\tikzstyle{state1a}=[ellipse,draw=black!75, node distance=1.2cm]
	\tikzstyle{state2}=[ellipse,draw=black!75, node distance=1.75cm]
	\tikzstyle{state3}=[ellipse,draw=black!75, node distance=2cm]
	\tikzstyle{state4}=[ellipse,draw=black!75, node distance=2.6cm]
	\tikzstyle{input}=[draw=none,node distance=1.2cm]
	\tikzstyle{output}=[draw=none,node distance=1cm]
	\tikzstyle{dummy1}=[draw=none,node distance=1.14cm]
	\tikzstyle{dummy2}=[draw=none,node distance=1.8cm]
	\tikzstyle{dummy3}=[draw=none,node distance=1cm]

	\node[state2](sigma2)[]			{$\quad^{ }$};	\draw (sigma2)+(0,0) node {$\sigma_2$};
	\node[state2](sigma3)[right of=sigma2]	{$\quad^{ }$};	\draw (sigma3)+(0,0) node {$\sigma_3$};
	\node[state2](sigma4)[right of=sigma3]	{$\quad^{ }$};	\draw (sigma4)+(0,0) node {$\sigma_4$};
	\node[state2](sigma5)[right of=sigma4]	{$\quad^{ }$};	\draw (sigma5)+(0,0) node {$\sigma_5$};
	\node[state2](sigma6)[right of=sigma5]	{$\quad^{ }$};	\draw (sigma6)+(0,0) node {$\sigma_6$};
	\node[state2](sigma7)[right of=sigma6]	{$\quad^{ }$};	\draw (sigma7)+(0,0) node {$\sigma_7$};
	\node[dummy1](dummy1)[above right of=sigma4]	{};
	\node[dummy1](dummy1a)[above of=dummy1]	{};
	\node[state2](sigma1)[above of=dummy1a]	{$\quad^{ }$};	\draw (sigma1)+(0,0) node {$\sigma_1$};
	\node[dummy1](dummy2)[below of=dummy1]	{};
	\node[dummy1](dummy3)[below of=dummy2]	{};
	\node[state2](sigma9)[below of=dummy3]{counter 2};	\draw (sigma9)+(-1.44,0) node {$\sigma_{9}$};
	\node[dummy2](dummy8)[left of=sigma9]	{};
	\node[state](sigma8)[left of=dummy8]{counter 1};	\draw (sigma8)+(-1.42,0) node {$\sigma_{8}$};
	\node[dummy2](dummy10)[right of=sigma9]	{};
	\node[state](sigma10)[right of=dummy10]{counter 3};	\draw (sigma10)+(-1.43,0) node {$\sigma_{10}$};
	\node[state4](sigma11)[right of=sigma10]	{$\quad^{ }$};	\draw (sigma11)+(0,0) node {$\sigma_{11}$};	
	\node[dummy3](dummy12)[right of=sigma1]	{};
	\node[dummy3](dummy12a)[below right of=dummy12]	{};
	\node[state3](sigma12)[right of=dummy12a]{$\quad^{ }$};	\draw (sigma12)+(0,0) node {$\sigma_{12}$};	
	\node[state1](sigma13)[right of=sigma12]{$\quad^{ }$};	\draw (sigma13)+(0,0) node {$\sigma_{13}$};	
	\node[state1](sigma14)[right of=sigma13]{$\quad^{ }$};	\draw (sigma14)+(0,0) node {$\sigma_{14}$};	
	\node[state1a](sigma15)[below of=sigma12]{$\quad^{ }$};	\draw (sigma15)+(0,0) node {$\sigma_{15}$};	
	\node[state1](sigma16)[right of=sigma15]{$\quad^{ }$};	\draw (sigma16)+(0,0) node {$\sigma_{16}$};
	\node[state1](sigma17)[right of=sigma16]{$\quad^{ }$};	\draw (sigma17)+(0,0) node {$\sigma_{17}$};
	\node[input](input)[above of=sigma1]	{};
	\draw (sigma1)+(-.45,.63) node {\small input};	
	\node[output](output)[below of=sigma11]	{};

\path[->]	(input)edge	node 	{} (sigma1)
		(sigma1)edge	node 	{}      (sigma2)
		(sigma1)edge	node 	{}      (sigma3)
		(sigma1)edge	node 	{}      (sigma4)
		(sigma1)edge	node	{}	(sigma5)
		(sigma1)edge	node	{}	(sigma6)
		(sigma1)edge	node	{}	(sigma7)
		(sigma1)edge	node	{}	(sigma8)
		(sigma1)edge	node	{}	(sigma9)
		(sigma1)edge	node	{}	(sigma10)
		(sigma1)edge	node	{}	(sigma12)
		(sigma1)edge	[bend left,in=160]node	{}	(sigma13)
		(sigma1)edge	[bend left,in=160]node	{}	(sigma14)
		(sigma2)edge	node 	{}      (sigma3)
		(sigma2)edge	[bend left,in=140]node {}      (sigma4)
		(sigma2)edge	[bend left,in=140]node {}      (sigma5)
		(sigma2)edge	[bend left,in=145]node {}      (sigma6)
		(sigma2)edge	[bend left,in=152]node {}      (sigma7)
		(sigma2)edge	node	{}	(sigma8)
		(sigma2)edge	node	{}	(sigma9)
		(sigma2)edge	node	{}	(sigma10)
		(sigma3)edge	node	{}	(sigma2)
		(sigma3)edge	node	{}	(sigma8)
		(sigma3)edge	node	{}	(sigma9)
		(sigma3)edge	node	{}	(sigma10)
		(sigma4)edge	node	{}	(sigma8)
		(sigma4)edge	node	{}	(sigma9)
		(sigma4)edge	node	{}	(sigma10)
		(sigma5)edge	node	{}	(sigma8)
		(sigma5)edge	node	{}	(sigma9)
		(sigma5)edge	node	{}	(sigma10)
		(sigma6)edge	node	{}	(sigma8)
		(sigma6)edge	node	{}	(sigma9)
		(sigma6)edge	node	{}	(sigma10)
		(sigma7)edge	node	{}	(sigma8)
		(sigma7)edge	node	{}	(sigma9)
		(sigma7)edge	node	{}	(sigma10)
		(sigma8)edge	node	{}	(sigma2)
		(sigma8)edge	node	{}	(sigma3)
		(sigma8)edge	node	{}	(sigma4)
		(sigma8)edge	node	{}	(sigma5)
		(sigma8)edge	node	{}	(sigma6)
		(sigma8)edge	node	{}	(sigma7)
		(sigma9)edge	node	{}	(sigma2)
		(sigma9)edge	node	{}	(sigma3)
		(sigma9)edge	node	{}	(sigma4)
		(sigma9)edge	node	{}	(sigma5)
		(sigma9)edge	node	{}	(sigma6)
		(sigma9)edge	node	{}	(sigma7)
		(sigma10)edge	node	{}	(sigma2)
		(sigma10)edge	node	{}	(sigma3)
		(sigma10)edge	node	{}	(sigma4)
		(sigma10)edge	node	{}	(sigma5)
		(sigma10)edge	node	{}	(sigma6)
		(sigma10)edge	node	{}	(sigma7)
		(sigma10)edge	node	{}	(sigma11)
		(sigma12)edge	node	{}	(sigma15)
		(sigma12)edge	node	{}	(sigma16)
		(sigma12)edge	node	{}	(sigma17)
		(sigma13)edge	node	{}	(sigma15)
		(sigma13)edge	node	{}	(sigma16)
		(sigma13)edge	node	{}	(sigma17)
		(sigma14)edge	node	{}	(sigma15)
		(sigma14)edge	node	{}	(sigma16)
		(sigma14)edge	node	{}	(sigma17)
		(sigma11)edge	node	{\small output}	(output);

	\end{tikzpicture}
            \end{center}
\caption{Part 1 of the universal SN P system $\Pi'_{C_3}$. Each oval labeled $\sigma_i$ is a neuron. An arrow going from neuron $\sigma_i$ to neuron $\sigma_j$ illustrates a synapse $(i,j)$.}\label{fig:small universal SNP system I}
\end{figure}

\begin{pf}
Let $C_3=(3,\{c_{1},c_{2}\},c_{3},Q,q_1,q_h)$ where $Q=\{q_1,q_2,\cdots,q_h\}$. Also, without loss of generality we assume that during $C_3$'s computation if $C_3$ is decrementing a counter no other counter has value 0 at that timestep (see the paragraph before Theorem~\ref{thm:universal SNP systems with 17 neuron}). The SN P system $\Pi'_{C_3}$ is given by Figures~\ref{fig:small universal SNP system I} and~\ref{fig:small universal SNP system II}  and Tables~\ref{tab:neurons of SNP I} and~\ref{tab:neurons of SNP II}. As a complement to the figures, Table~\ref{tab:synapses} may be used to identify all the synapses in $\Pi'_{C_3}$. The algorithm given for $\Pi'_{C_3}$ is deterministic.

\subsubsection{Encoding of a configuration of $C_3$ and reading input into $\Pi'_{C_3}$}\label{sec:Encoding of a configuration of C_3 and reading in input to Pi'_(C_3)}
A configuration of $C_3$ is stored as spikes in the neurons of $\Pi'_{C_3}$. The next instruction $q_i$ to be executed is stored in each of the neurons $\sigma_{2}$, $\sigma_{3}$, $\sigma_{4}$, $\sigma_{5}$, $\sigma_{6}$, and $\sigma_{7}$ as $21(h+i)+1$ spikes. Let $x_1$, $x_2$ and $x_3$ be the values stored in counters $c_1$, $c_2$ and $c_3$, respectively. Then the value $x_1$ is stored as $6(x_1+1)$ spikes in neuron $\sigma_{8}$, $x_2$ is stored as $6(x_2+1)$ spikes in $\sigma_{9}$, and $x_3$ is stored as $6(x_3+1)$ spikes in $\sigma_{10}$.

The input to $\Pi'_{C_3}$ is read into the system via the input neuron $\sigma_1$ (see Figure~\ref{fig:small universal SNP system I}). If $C_3$ begins its computation with the values $x_1$ and $x_2$ in counters $c_1$ and $c_2$, respectively, then the binary sequence $w=10^{x_1-1}10^{x_2-1}1$ is read in via the input neuron $\sigma_1$. Thus, $\sigma_1$ receives a spike from the environment at times $t_1$, $t_{x_1+1}$ and $t_{x_1+x_2+1}$. We explain how the system is initialised to encode an initial configuration of $C_3$ by giving the number of spikes in each neuron and the rule that is to be applied in each neuron at time $t$. Before the computation begins neurons $\sigma_{2}$, $\sigma_{3}$, $\sigma_{4}$, $\sigma_{5}$, $\sigma_{6}$ and $\sigma_{7}$ each contain $40$ spikes, neurons $\sigma_{8}$, $\sigma_{9}$ and $\sigma_{10}$ each contain $3$ spikes, and neurons $\sigma_{12}$, $\sigma_{13}$ and $\sigma_{14}$ each contain $21h-2$ spikes. Thus, when $\sigma_1$ receives it first spike at time $t_1$ we have
\begin{xalignat*}{2}
&t_{1}:\; \\
	&\qquad\sigma_1=1,	& s\rightarrow s, \\
	&\qquad\sigma_{2},\sigma_{3},\sigma_{4},\sigma_{5},\sigma_{6},\sigma_{7}=40,&\\
	&\qquad\sigma_{8},\sigma_{9},\sigma_{10}=3, &\\
	&\qquad\sigma_{12},\sigma_{13},\sigma_{14}=21h-2, 	& (s^3)^{\ast}s^4/s^3\rightarrow s. 
\end{xalignat*}
Thus, from Figures~\ref{fig:small universal SNP system I} and~\ref{fig:small universal SNP system II}, when we apply the rule $s\rightarrow s$ in neuron $\sigma_1$ and the rule $(s^3)^{\ast}s^4/s^3\rightarrow s$ in $\sigma_{12}$, $\sigma_{13}$ and $\sigma_{14}$ at time $t_1$ we get
\begin{xalignat*}{2}
&t_{2}:\; \\
	&\qquad\sigma_{2},\sigma_{3},\sigma_{4},\sigma_{5},\sigma_{6},\sigma_{7}=41,	& s^{41}/s\rightarrow s,\\
	&\qquad\sigma_{8},\sigma_{9},\sigma_{10}=4, &\\
	&\qquad\sigma_{12},\sigma_{13},\sigma_{14}=21h-4, &\\
	&\qquad\sigma_{15},\sigma_{16},\sigma_{17}=3, &\\
\\
&t_{3}:\; \\
	&\qquad\sigma_{2},\sigma_{3},\sigma_{4},\sigma_{5},\sigma_{6},\sigma_{7}=41,	& s^{41}/s\rightarrow s,\\
	&\qquad\sigma_{8}=10, &\\
	&\qquad\sigma_{9},\sigma_{10}=10, &(s^{6})^{\ast}s^{10}/s^{6}\rightarrow s,\\
	&\qquad\sigma_{11}=6,	& s^6\rightarrow \lambda,\\
	&\qquad\sigma_{12},\sigma_{13},\sigma_{14}=21h-4, &\\
	&\qquad\sigma_{15},\sigma_{16},\sigma_{17}=3. &
\end{xalignat*}
\begin{xalignat*}{2}
&t_{4}:\; \\
	&\qquad\sigma_{2},\sigma_{3},\sigma_{4},\sigma_{5},\sigma_{6},\sigma_{7}=43,	& s^{43}/s^3\rightarrow s,\\
	&\qquad\sigma_{8}=16, &\\
	&\qquad\sigma_{9},\sigma_{10}=10, &(s^{6})^{\ast}s^{10}/s^{6}\rightarrow s,\\
	&\qquad\sigma_{11}=7,	& s^7\rightarrow \lambda,\\
	&\qquad\sigma_{12},\sigma_{13},\sigma_{14}=21h-4, &\\
	&\qquad\sigma_{15},\sigma_{16},\sigma_{17}=3. &
\end{xalignat*}
Neurons $\sigma_{2}$, $\sigma_{3}$, $\sigma_{4}$, $\sigma_{5}$, $\sigma_{6}$ and $\sigma_{7}$ fire on every timestep between times $t_2$ and $t_{x_1+2}$ to send a total of $6x_1$ spikes to $\sigma_{8}$, and thus we get
\begin{xalignat*}{2}
&t_{x_1+1}:\; \\
	&\qquad\sigma_1=1,	& s\rightarrow s, \\
	&\qquad\sigma_{2},\sigma_{3},\sigma_{4},\sigma_{5},\sigma_{6},\sigma_{7}=43,	& s^{43}/s^3\rightarrow s,\\
	&\qquad\sigma_{8}=6(x_1-1)+4, &\\
	&\qquad\sigma_{9},\sigma_{10}=10, &(s^{6})^{\ast}s^{10}/s^{6}\rightarrow s,\\
	&\qquad\sigma_{11}=7,	& s^7\rightarrow \lambda,\\
	&\qquad\sigma_{12},\sigma_{13},\sigma_{14}=21h-4, &\\
	&\qquad\sigma_{15},\sigma_{16},\sigma_{17}=3, &\\
\\
&t_{x_1+2}:\; \\
	&\qquad\sigma_{2}=44,	& s^{44}/s^{25}\rightarrow s,\\
	&\qquad\sigma_{3},\sigma_{4},\sigma_{5},\sigma_{6},\sigma_{7}=44,	& s^{44}/s^{31}\rightarrow s,\\
	&\qquad\sigma_{8}=6x_1+5, &(s^{6})^{\ast}s^{11}/s^6\rightarrow s,\\
	&\qquad\sigma_{9}=11, &\\
	&\qquad\sigma_{10}=11, &(s^{6})^{\ast}s^{11}/s^{6}\rightarrow s,\\
	&\qquad\sigma_{11}=7,	& s^7\rightarrow \lambda,\\
	&\qquad\sigma_{12},\sigma_{13},\sigma_{14}=21h-3,&\\
	&\qquad\sigma_{15},\sigma_{16},\sigma_{17}=3, &\\
\\
&t_{x_1+3}:\; \\
	&\qquad\sigma_{2}=22,	& s^{22}/s^3\rightarrow s,\\
	&\qquad\sigma_{3},\sigma_{4},\sigma_{5},\sigma_{6},\sigma_{7}=16,	& s^{16}/s^3\rightarrow s,\\
	&\qquad\sigma_{8}=6x_1+5, &(s^{6})^{\ast}s^{11}/s^6\rightarrow s,\\
	&\qquad\sigma_{9}=17, &\\
	&\qquad\sigma_{10}=11, &(s^{6})^{\ast}s^{11}/s^{6}\rightarrow s,\\
	&\qquad\sigma_{11}=7,	& s^7\rightarrow \lambda,\\
	&\qquad\sigma_{12},\sigma_{13},\sigma_{14}=21h-3,&\\
	&\qquad\sigma_{15},\sigma_{16},\sigma_{17}=3.
\end{xalignat*}
Neurons $\sigma_{2}$, $\sigma_{3}$, $\sigma_{4}$, $\sigma_{5}$, $\sigma_{6}$ and $\sigma_{7}$ fire on every timestep between times $t_{x_1+2}$ and $t_{x_1+x_2+2}$ to send a total of $6x_2$ spikes to $\sigma_{9}$. Thus, when $\sigma_1$ receives the last spike from its environment we have
\begin{xalignat*}{2}
&t_{x_1+x_2+1}:\; \\
	&\qquad\sigma_1=1,	& s\rightarrow s, \\
	&\qquad\sigma_{2}=22,	& s^{22}/s^3\rightarrow s,\\
	&\qquad\sigma_{3},\sigma_{4},\sigma_{5},\sigma_{6},\sigma_{7}=16,	& s^{16}/s^3\rightarrow s,\\
	&\qquad\sigma_{8}=6x_1+5, &(s^{6})^{\ast}s^{11}/s^6\rightarrow s,\\
	&\qquad\sigma_{9}=6x_2+5, &\\
	&\qquad\sigma_{10}=11, &(s^{6})^{\ast}s^{11}/s^{6}\rightarrow s,\\
	&\qquad\sigma_{11}=7,	& s^7\rightarrow \lambda,\\
	&\qquad\sigma_{12},\sigma_{13},\sigma_{14}=21h-3,&\\
	&\qquad\sigma_{15},\sigma_{16},\sigma_{17}=3,	&\\
\\
&t_{x_1+x_2+2}:\; \\
	&\qquad\sigma_{2}=23,	& s^{23}/s^{5}\rightarrow s,\\
	&\qquad\sigma_{3},\sigma_{4},\sigma_{5},\sigma_{6},\sigma_{7}=17,	& \\
	&\qquad\sigma_{8}=6(x_1+1), &\\
	&\qquad\sigma_{9}=6(x_2+2), &\\
	&\qquad\sigma_{10}=12, &\\
	&\qquad\sigma_{11}=7,	& s^7\rightarrow \lambda,\\
	&\qquad\sigma_{12},\sigma_{13},\sigma_{14}=21h-2, &(s^{3})^{\ast}s^4/s^3\rightarrow s,\\
	&\qquad\sigma_{15},\sigma_{16},\sigma_{17}=3,	&\\
\\
&t_{x_1+x_2+3}:\; \\
	&\qquad\sigma_{2},\sigma_{3},\sigma_{4},\sigma_{5},\sigma_{6},\sigma_{7}=18,	& \\
	&\qquad\sigma_{8}=6(x_1+1)+1, &(s^{6})^{\ast}s^{13}/s\rightarrow s,\\
	&\qquad\sigma_{9}=6(x_2+2)+1, &(s^{6})^{\ast}s^{7}/s^{7}\rightarrow s,\\
	&\qquad\sigma_{10}=13, &(s^{6})^{\ast}s^{7}/s^{7}\rightarrow s,\\
	&\qquad\sigma_{11}=1,	& s\rightarrow \lambda,\\
	&\qquad\sigma_{12},\sigma_{13},\sigma_{14}=21h-5, &(s^{3})^{\ast}s^4/s^3\rightarrow s,\\
	&\qquad\sigma_{15},\sigma_{16},\sigma_{17}=6,	&\\
&t_{x_1+x_2+4}:\; \\
	&\qquad\sigma_{2},\sigma_{3},\sigma_{4},\sigma_{5},\sigma_{6},\sigma_{7}=21,	& \\
	&\qquad\sigma_{8}=6(x_1+1), &\\
	&\qquad\sigma_{9}=6(x_2+1), &\\
	&\qquad\sigma_{10}=6, &\\
	&\qquad\sigma_{11}=1,	& s\rightarrow \lambda,\\
	&\qquad\sigma_{12},\sigma_{13},\sigma_{14}=21h-8, &(s^{3})^{\ast}s^4/s^3\rightarrow s,\\
	&\qquad\sigma_{15},\sigma_{16},\sigma_{17}=9.
\end{xalignat*}\\

After a further $7h-3$ timestep we get
\begin{xalignat*}{2}
&t_{x_1+x_2+7h+1}:\; \\
	&\qquad\sigma_{2},\sigma_{3},\sigma_{4},\sigma_{5},\sigma_{6},\sigma_{7}=21,	& \\
	&\qquad\sigma_{8}=6(x_1+1), &\\
	&\qquad\sigma_{9}=6(x_2+1), &\\
	&\qquad\sigma_{10}=6, &\\
	&\qquad\sigma_{12}=1, &s\rightarrow s,\\
	&\qquad\sigma_{13},\sigma_{14}=1, &s\rightarrow \lambda,\\
	&\qquad\sigma_{15},\sigma_{16},\sigma_{17}=21h,\\
\\
&t_{x_1+x_2+7h+2}:\; \\
	&\qquad\sigma_{2},\sigma_{3},\sigma_{4},\sigma_{5},\sigma_{6},\sigma_{7}=21,	& \\
	&\qquad\sigma_{8}=6(x_1+1), &\\
	&\qquad\sigma_{9}=6(x_2+1), &\\
	&\qquad\sigma_{10}=6, &\\
	&\qquad\sigma_{15},\sigma_{16},\sigma_{17}=21h+1, &(s^{3})^{\ast}s^4/s^3\rightarrow s,\\
\\
&t_{x_1+x_2+7h+3}:\; \\
	&\qquad\sigma_{2},\sigma_{3},\sigma_{4},\sigma_{5},\sigma_{6},\sigma_{7}=21+3,	& \\
	&\qquad\sigma_{8}=6(x_1+1), &\\
	&\qquad\sigma_{9}=6(x_2+1), &\\
	&\qquad\sigma_{10}=6, &\\
	&\qquad\sigma_{12},\sigma_{13},\sigma_{14}=3, &\\
	&\qquad\sigma_{15},\sigma_{16},\sigma_{17}=21h-2, &(s^{3})^{\ast}s^4/s^3\rightarrow s.
\end{xalignat*}
Neurons $\sigma_{15}$, $\sigma_{16}$ and $\sigma_{17}$ continue to fire at each timestep. Thus, after a further $7h-1$ steps we get
\begin{xalignat*}{2}
&t_{x_1+x_2+14h+2}:\; \\
	&\qquad\sigma_{2},\sigma_{3},\sigma_{4},\sigma_{5},\sigma_{6},\sigma_{7}=21h+21,	& \\
	&\qquad\sigma_{8}=6(x_1+1), &\\
	&\qquad\sigma_{9}=6(x_2+1), &\\
	&\qquad\sigma_{10}=6, &\\
	&\qquad\sigma_{12},\sigma_{13},\sigma_{14}=21h, &\\
	&\qquad\sigma_{15}=1, &s\rightarrow s,\\
	&\qquad\sigma_{16},\sigma_{17}=1, &s\rightarrow \lambda.
\end{xalignat*}
\begin{xalignat*}{2}
&t_{x_1+x_2+14h+3}:\; \\
	&\qquad\sigma_{2},\sigma_{3},\sigma_{4},\sigma_{5},\sigma_{6},\sigma_{7}=21(h+1)+1,	& \\
	&\qquad\sigma_{8}=6(x_1+1), &\\
	&\qquad\sigma_{9}=6(x_2+1), &\\
	&\qquad\sigma_{10}=6, &\\
	&\qquad\sigma_{12},\sigma_{13},\sigma_{14}=21h+1.
\end{xalignat*}
At time $t_{x_1+x_2+14h+3}$ neurons  $\sigma_{2}$, $\sigma_{3}$, $\sigma_{4}$, $\sigma_{5}$, $\sigma_{6}$ and $\sigma_{7}$ each contain $21(h+1)+1$ spikes, $\sigma_8$ contains $6(x_1+1)$ spikes, $\sigma_{9}$ contains $6(x_2+1)$ spikes and $\sigma_{10}$ contains $6$ spikes. Thus, at time $t_{x_1+x_2+14h+3}$ the SN P system encodes an initial configuration of $C_3$. 
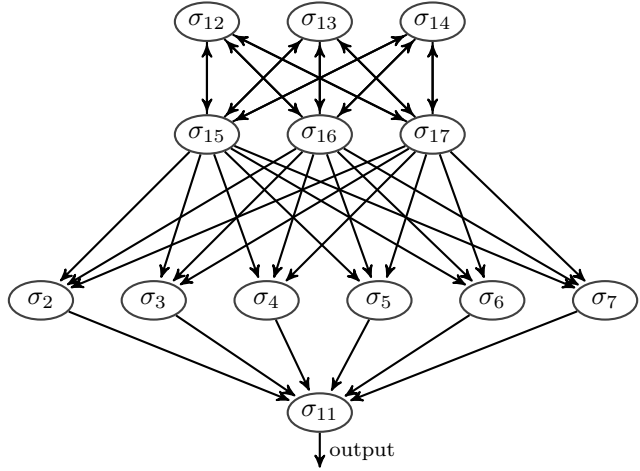
\begin{figure}[t]
\begin{center}
\begin{tikzpicture}[>=stealth',shorten >=1pt,auto,node distance=1.5cm,thick,bend angle=45]

	\tikzstyle{dots}=[draw=none,node distance=1cm]
	\tikzstyle{state2}=[ellipse,draw=black!75, node distance=1.5cm]
	\tikzstyle{state3}=[ellipse,draw=black!75, node distance=1.2cm]
	\tikzstyle{dummy3}=[draw=none,node distance=1cm]
	\tikzstyle{output}=[draw=none,node distance=.9cm]

	\node[state2](sigma2)[]			{$\quad^{ }$};	\draw (sigma2)+(0,0) node {$\sigma_2$};
	\node[state2](sigma3)[right of=sigma2]	{$\quad^{ }$};	\draw (sigma3)+(0,0) node {$\sigma_3$};
	\node[state2](sigma4)[right of=sigma3]	{$\quad^{ }$};	\draw (sigma4)+(0,0) node {$\sigma_4$};
	\node[state2](sigma5)[right of=sigma4]	{$\quad^{ }$};	\draw (sigma5)+(0,0) node {$\sigma_5$};
	\node[state2](sigma6)[right of=sigma5]	{$\quad^{ }$};	\draw (sigma6)+(0,0) node {$\sigma_6$};
	\node[state2](sigma7)[right of=sigma6]	{$\quad^{ }$};	\draw (sigma7)+(0,0) node {$\sigma_7$};
	\node[dummy3](dummy16)[above right of=sigma4]	{};
	\node[state2](sigma16)[above of=dummy16]{$\quad^{ }$};	\draw (sigma16)+(0,0) node {$\sigma_{16}$};
	\node[state2](sigma15)[left of=sigma16]{$\quad^{ }$};	\draw (sigma15)+(0,0) node {$\sigma_{15}$};	
	\node[state2](sigma17)[right of=sigma16]{$\quad^{ }$};	\draw (sigma17)+(0,0) node {$\sigma_{17}$};
	\node[state2](sigma12)[above of=sigma15]{$\quad^{ }$};	\draw (sigma12)+(0,0) node {$\sigma_{12}$};	
	\node[state2](sigma13)[right of=sigma12]{$\quad^{ }$};	\draw (sigma13)+(0,0) node {$\sigma_{13}$};	
	\node[state2](sigma14)[right of=sigma13]{$\quad^{ }$};	\draw (sigma14)+(0,0) node {$\sigma_{14}$};	
	\node[dummy3](dummy11)[below of=dummy16]	{};
	\node[state3](sigma11)[below of=dummy11]{$\quad^{ }$};	\draw (sigma11)+(0,0) node {$\sigma_{11}$};	
	\node[output](output)[below of=sigma11]	{};

\path[->]	(sigma2)edge	node	{}	(sigma11)
		(sigma3)edge	node	{}	(sigma11)
		(sigma4)edge	node	{}	(sigma11)
		(sigma5)edge	node	{}	(sigma11)
		(sigma6)edge	node	{}	(sigma11)
		(sigma7)edge	node	{}	(sigma11)
		(sigma12)edge	node	{}	(sigma15)
		(sigma12)edge	node	{}	(sigma16)
		(sigma12)edge	node	{}	(sigma17)
		(sigma13)edge	node	{}	(sigma15)
		(sigma13)edge	node	{}	(sigma16)
		(sigma13)edge	node	{}	(sigma17)
		(sigma14)edge	node	{}	(sigma15)
		(sigma14)edge	node	{}	(sigma16)
		(sigma14)edge	node	{}	(sigma17)
		(sigma15)edge	node	{}	(sigma2)
		(sigma15)edge	node	{}	(sigma3)
		(sigma15)edge	node	{}	(sigma4)
		(sigma15)edge	node	{}	(sigma5)
		(sigma15)edge	node	{}	(sigma6)
		(sigma15)edge	node	{}	(sigma7)
		(sigma15)edge	node	{}	(sigma12)
		(sigma15)edge	node	{}	(sigma13)
		(sigma15)edge	node	{}	(sigma14)
		(sigma16)edge	node	{}	(sigma2)
		(sigma16)edge	node	{}	(sigma3)
		(sigma16)edge	node	{}	(sigma4)
		(sigma16)edge	node	{}	(sigma5)
		(sigma16)edge	node	{}	(sigma6)
		(sigma16)edge	node	{}	(sigma7)
		(sigma16)edge	node	{}	(sigma12)
		(sigma16)edge	node	{}	(sigma13)
		(sigma16)edge	node	{}	(sigma14)
		(sigma17)edge	node	{}	(sigma2)
		(sigma17)edge	node	{}	(sigma3)
		(sigma17)edge	node	{}	(sigma4)
		(sigma17)edge	node	{}	(sigma5)
		(sigma17)edge	node	{}	(sigma6)
		(sigma17)edge	node	{}	(sigma7)
		(sigma17)edge	node	{}	(sigma12)
		(sigma17)edge	node	{}	(sigma13)
		(sigma17)edge	node	{}	(sigma14)
		(sigma11)edge	node	{\small output}	(output);

	\end{tikzpicture}
            \end{center}
\caption{Part 2 of the universal SN P system $\Pi'_{C_3}$. Each oval labeled $\sigma_i$ is a neuron. An arrow going from neuron $\sigma_i$ to neuron $\sigma_j$ illustrates a synapse $(i,j)$.}\label{fig:small universal SNP system II}
\end{figure}

\subsubsection{Algorithm overview}\label{sec:Algorithm overview Pi'_(C_3)}
Here we give a high level overview of the simulation algorithm used by $\Pi'_{C_3}$. Neurons $\sigma_{8}$, $\sigma_{9}$ and $\sigma_{10}$ simulate the counters of $c_1$, $c_2$ and $c_3$, respectively. Neurons $\sigma_{2}$, $\sigma_{3}$, $\sigma_{4}$, $\sigma_{5}$, $\sigma_{6}$ and $\sigma_{7}$ are the \emph{control neurons}. They determine which instruction is to be simulated next by sending signals to the neurons that simulate the counters of $C_3$ directing them to simulate an increment or decrement. There are four different signals that the control neurons send to the simulated counters. Each of these signals takes the form of a unique number of spikes. If 1 spike is sent to $\sigma_{8}$, $\sigma_{9}$ and $\sigma_{10}$ then the value in $\sigma_{8}$ (counter $c_1$) is tested and $\sigma_{9}$ (counter $c_2$) and $\sigma_{10}$ (counter $c_3$) are decremented. If 2 spikes are sent the value of $\sigma_{9}$ is tested and $\sigma_{8}$ and $\sigma_{10}$ are decremented. If 3 spikes are sent the value of $\sigma_{10}$ is tested and $\sigma_{8}$ and $\sigma_{9}$ are decremented. Finally, if 6 spikes are sent all three counters are incremented. Unfortunately, all of the above signals have the effect of changing the value of more than one simulated counter at a time. We can, however, obtain the desired result by using more than one signal for each simulated timestep. If we wish to simulate $INC$ we send 2 signals and if we wish to simulate $DEC$ we send either 8 or 2 signals. Table~\ref{tab:simulate counter instruction} gives the sequence of spikes (signals) to be sent in order to simulate each counter machine instruction. To explain how to use Table~\ref{tab:simulate counter instruction} we will take the example of simulating $INC(2)$. In the first timestep, all three simulated counters $\sigma_{8}$, $\sigma_{9}$ and $\sigma_{10}$ are incremented by sending 6 spikes, and then in the second timestep the simulated counters $\sigma_{8}$ and $\sigma_{10}$ are decremented by sending 2 spikes. This has the effect of simulating an increment in counter $c_2$ and leaving the other two simulated counters unchanged. 

Each counter machine instruction $q_i$ is encoded as $21(h+i)+1$ spikes in each of the control neurons. At the end of each simulated timestep the number of spikes in the control neurons must be updated to encode the next instruction~$q_k$. The update rule $s^{21(h+i)-21k}\rightarrow s$ is applied in each control neuron leaving a total of $21k$ spikes in each control neuron. Following this, $21h+1$ spikes are sent from neurons $\sigma_{15}$, $\sigma_{16}$ and $\sigma_{17}$ to each of the control neurons. This gives a total of $21(h+k)+1$ spikes in each control neuron. Thus encoding the next instruction $q_k$. (Note that the rule $s^{21(h+i)-21k}\rightarrow s$ is simplification of the actual rule used.)
\begin{table}[t]
\begin{center}
\begin{tabular}{c@{\;\;\;}|@{\;\;\;}l}
Instruction	& Sequence of spikes sent from $\sigma_{2}$, $\sigma_{3}$, $\sigma_{4}$, $\sigma_{5}$, $\sigma_{6}$ and $\sigma_{7}$\\ \hline
$INC(1)$& 6, 1\\
$INC(2)$& 6, 2\\
$INC(3)$& 6, 3\\
$DEC(1)$& 1, 0, 6 if $x_1=0$\\
$DEC(1)$& 1, 0, 6, 6, 6, 3, 3, 2, 2 if $x_1>0$\\
$DEC(2)$& 2, 0, 6 if $x_2=0$\\
$DEC(2)$& 2, 0, 6, 6, 6, 3, 3, 1, 1 if $x_2>0$\\
$DEC(3)$& 3, 0, 6 if $x_3=0$\\
$DEC(3)$& 3, 0, 6, 6, 6, 2, 2, 1, 1 if $x_3>0$\\
\hline
\end{tabular}
\end{center}
\caption{This table gives a counter machine instruction in the left column followed, in the right column, by the sequence that is used by $\Pi'_{C_3}$ to simulated that instruction. Each number in the sequence represents the total number of spikes to be sent from the set of neurons $\sigma_{2}$, $\sigma_{3}$, $\sigma_{4}$, $\sigma_{5}$, $\sigma_{6}$ and $\sigma_{7}$ at each timestep.}\label{tab:simulate counter instruction}
\end{table}

\subsubsection{$\Pi'_{C_3}$ simulating $q_i:INC(1),q_l$}
The simulation of $INC(1)$ is given by the neurons in Figures~\ref{fig:small universal SNP system I} and~\ref{fig:small universal SNP system II}. Let $x_1$, $x_2$ and $x_3$ be the values in counters $c_1$, $c_2$ and $c_3$ respectively. Then our simulation of $q_i:INC(1),q_l$ begins with $6(x_1+1)$ spikes in $\sigma_{8}$, $6(x_2+1)$ spikes in $\sigma_{9}$, $6(x_3+1)$ spikes in $\sigma_{10}$, $21(h+i)+1$ spikes in each of the neurons $\sigma_{2}$, $\sigma_{3}$, $\sigma_{4}$, $\sigma_{5}$, $\sigma_{6}$ and $\sigma_{7}$, and $21h+1$ spikes in each of the neurons $\sigma_{12}$, $\sigma_{13}$ and $\sigma_{14}$. Beginning our simulation at time $t_j$, we have
\begin{xalignat*}{2}
&t_{j}:\; \\
	&\qquad\sigma_{2}=21(h+i)+1,	&s^{21(h+i)+1}/s^4\rightarrow s,\\
	&\qquad\sigma_{3},\sigma_{4},\sigma_{5},\sigma_{6},\sigma_{7}=21(h+i)+1, &s^{21(h+i)+1}/s^{21(h+i-l)+6}\rightarrow s,\\
	&\qquad\sigma_{8}=6(x_1+1), &\\
	&\qquad\sigma_{9}=6(x_2+1), &\\
	&\qquad\sigma_{10}=6(x_3+1), &\\
	&\qquad\sigma_{12},\sigma_{13},\sigma_{14}=21h+1, &(s^{3})^{\ast}s^4/s^3\rightarrow s.
\end{xalignat*}\newpage
Thus, from Figures~\ref{fig:small universal SNP system I} and~\ref{fig:small universal SNP system II} we get
\begin{xalignat*}{2}
&t_{j+1}:\; \\
	&\qquad\sigma_{2}=21(h+i)-2,	&s^{21(h+i)-2}/s^{21(h+i-l)+1}\rightarrow s,\\
	&\qquad\sigma_{3},\sigma_{4},\sigma_{5},\sigma_{6},\sigma_{7}=21l-4, &\\
	&\qquad\sigma_{8}=6(x_1+2), &\\
	&\qquad\sigma_{9}=6(x_2+2), &\\
	&\qquad\sigma_{10}=6(x_3+2), &\\
	&\qquad\sigma_{11}=6, & s^6\rightarrow \lambda,\\
	&\qquad\sigma_{12},\sigma_{13},\sigma_{14}=21h-2, &(s^{3})^{\ast}s^4/s^3\rightarrow s,\\
	&\qquad\sigma_{15},\sigma_{16},\sigma_{17}=3, &\\
\\
&t_{j+2}:\; \\
	&\qquad\sigma_{2},\sigma_{3},\sigma_{4},\sigma_{5},\sigma_{6},\sigma_{7}=21l-3, &\\
	&\qquad\sigma_{8}=6(x_1+2)+1, &(s^{6})^{\ast}s^{13}/s\rightarrow s,\\
	&\qquad\sigma_{9}=6(x_2+2)+1, &(s^{6})^{\ast}s^{7}/s^{7}\rightarrow s,\\
	&\qquad\sigma_{10}=6(x_3+2)+1, &(s^{6})^{\ast}s^{7}/s^{7}\rightarrow s,\\
	&\qquad\sigma_{11}=1, & s\rightarrow \lambda,\\
	&\qquad\sigma_{12},\sigma_{13},\sigma_{14}=21h-5, &(s^{3})^{\ast}s^4/s^3\rightarrow s,\\
	&\qquad\sigma_{15},\sigma_{16},\sigma_{17}=6, &\\
\\
&t_{j+3}:\; \\
	&\qquad\sigma_{2},\sigma_{3},\sigma_{4},\sigma_{5},\sigma_{6},\sigma_{7}=21l, &\\
	&\qquad\sigma_{8}=6(x_1+2), &\\
	&\qquad\sigma_{9}=6(x_2+1), &\\
	&\qquad\sigma_{10}=6(x_3+1),&\\
	&\qquad\sigma_{11}=1, & s\rightarrow \lambda,\\
	&\qquad\sigma_{12},\sigma_{13},\sigma_{14}=21h-8, &(s^{3})^{\ast}s^4/s^3\rightarrow s,\\
	&\qquad\sigma_{15},\sigma_{16},\sigma_{17}=9. &\\
\end{xalignat*}
The remainder of this simulation is similar to the computation carried out at the end of the initialisation process (see the last paragraph of Section~\ref{sec:Algorithm overview Pi'_(C_3)} and timesteps $t_{x_1+x_2+4}$ to $t_{x_1+x_2+14h+3}$ of the Section~\ref{sec:Encoding of a configuration of C_3 and reading in input to Pi'_(C_3)}). Thus, after a further $14h-1$ timesteps we get
\begin{xalignat*}{2}
&t_{j+14h+2}:\; \\
	&\qquad\sigma_{2},\sigma_{3},\sigma_{4},\sigma_{5},\sigma_{6},\sigma_{7}=21(h+l)+1, &\\
	&\qquad\sigma_{8}=6(x_1+2), &\\
	&\qquad\sigma_{9}=6(x_2+1), &\\
	&\qquad\sigma_{10}=6(x_3+1),&\\
	&\qquad\sigma_{12},\sigma_{13},\sigma_{14}=21h+1, &(s^{3})^{\ast}s^4/s^3\rightarrow s.\\
\end{xalignat*}
At time $t_{j+14h+2}$ the simulation of $q_i:INC(1),q_l$ is complete. Note that an increment on the value $x_1$ in counter $c_1$ is simulated by increasing the number of spikes in $\sigma_{8}$ from $6(x_1+1)$ to $6(x_1+2)$. Note also that the encoding of the next instruction $q_{l}$ is given by the $21(h+l)+1$ spikes in neurons $\sigma_{2}$, $\sigma_{3}$, $\sigma_{4}$, $\sigma_{5}$, $\sigma_{6}$ and $\sigma_{7}$.

\subsubsection{$\Pi'_{C_3}$ simulating $q_i:DEC(1),q_l,q_k$}
If we are simulating $DEC(1)$ then we get
\begin{xalignat*}{2}
&t_{j}:\; \\
	&\qquad\sigma_{2}=21(h+i)+1,	&s^{21(h+i)+1}/s^5\rightarrow s,\\
	&\qquad\sigma_{3},\sigma_{4},\sigma_{5},\sigma_{6},\sigma_{7}=21(h+i)+1, &\\
	&\qquad\sigma_{8}=6(x_1+1), &\\
	&\qquad\sigma_{9}=6(x_2+1), &\\
	&\qquad\sigma_{10}=6(x_3+1),&\\
	&\qquad\sigma_{12},\sigma_{13},\sigma_{14}=21h+1, &(s^{3})^{\ast}s^4/s^3\rightarrow s.\\ 
\end{xalignat*}
To help simplify configurations we will not include neurons $\sigma_{12}$, $\sigma_{13}$, and $\sigma_{14}$ until the end of the example. When simulating $DEC(1)$ there are two cases to consider. Case 1: if counter $c_1$ has value $x_1>0$, then decrement counter 1 and move to instruction $q_{i+1}$. Case 2: if counter $c_1$ has value $x_1=0$, then move to instruction $q_{k}$. In configuration $t_{j+1}$ our system determines if the value $x_1$ in counter 1 is $>0$ by checking if the number of spikes in $\sigma_{8}$ is $>13$. Note that if we have Case 1 then the rule $(s^6)^{\ast}s^{13}/s\rightarrow s$ is applied in $\sigma_{8}$ sending an extra spike to neurons $\sigma_{2}$, $\sigma_{3}$, $\sigma_{4}$, $\sigma_{5}$, $\sigma_{6}$ and $\sigma_{7}$ thus recording that $x_1>0$. Case 1 proceeds as follows:
\begin{xalignat*}{2}
&t_{j+1}:\; \\
	&\qquad\sigma_{2}=21(h+i)-4,	&\\
	&\qquad\sigma_{3},\sigma_{4},\sigma_{5},\sigma_{6},\sigma_{7}=21(h+i)+2, &\\
	&\qquad\sigma_{8}=6(x_1+1)+1, &(s^6)^{\ast}s^{13}/s\rightarrow s,\\
	&\qquad\sigma_{9}=6(x_2+1)+1, &(s^6)^{\ast}s^7/s^7\rightarrow s,\\
	&\qquad\sigma_{10}=6(x_3+1)+1, &(s^6)^{\ast}s^7/s^7\rightarrow s,\\ 
	&\qquad\sigma_{11}=1, & s\rightarrow \lambda,\\
\\
&t_{j+2}:\; \\
	&\qquad\sigma_{2}=21(h+i)-1,	&s^{21(h+i)-1}/s^5\rightarrow s,\\
	&\qquad\sigma_{3},\sigma_{4},\sigma_{5},\sigma_{6},\sigma_{7}=21(h+i)+5, &s^{21(h+i)+5}/s^{11}\rightarrow s,\\
	&\qquad\sigma_{8}=6(x_1+1), &\\
	&\qquad\sigma_{9}=6x_2, &\\
	&\qquad\sigma_{10}=6x_3&\\
	&\qquad\sigma_{11}=1, & s\rightarrow \lambda.
\end{xalignat*}
The method we use to test the value of $\sigma_{8}$ (simulated counter $c_1$) has the side-effect of decrementing $\sigma_{9}$ (simulated counter $c_2$) and $\sigma_{10}$ (simulated counter $c_2$). Following this, in order to get the correct values our algorithm takes the following steps: Each of our simulated counters ($\sigma_{8}$, $\sigma_{9}$ and $\sigma_{10}$) are incremented 3 times, and then the simulated counter $\sigma_{8}$ is decremented 4 times, whilst the simulated counters $\sigma_{9}$ and $\sigma_{10}$ are each decremented twice. Thus, the overall result is that a decrement of $c_1$ is simulated in $\sigma_{8}$ and the other encoded counter values in $\sigma_{9}$ and $\sigma_{10}$ remain the same. Continuing with our simulation we get
\begin{xalignat*}{2}
&t_{j+3}:\; \\
	&\qquad\sigma_{2},\sigma_{3},\sigma_{4},\sigma_{5},\sigma_{6},\sigma_{7}=21(h+i)-5,	&s^{21(h+i)-5}/s^3\rightarrow s,\\
	&\qquad\sigma_{8}=6(x_1+2), &\\
	&\qquad\sigma_{9}=6(x_2+1), &\\
	&\qquad\sigma_{10}=6(x_3+1), &\\
	&\qquad\sigma_{11}=6, & s^6\rightarrow \lambda, &\\
\\
&t_{j+4}:\; \\
	&\qquad\sigma_{2},\sigma_{3},\sigma_{4}=21(h+i)-7,	&s^{21(h+i)-7}/s^2\rightarrow s,\\
	&\qquad\sigma_{5},\sigma_{6},\sigma_{7}=21(h+i)-7,	&s^{21(h+i)-7}/s^{21(h+i-l)+10}\rightarrow s,\\
	&\qquad\sigma_{8}=6(x_1+3), &\\
	&\qquad\sigma_{9}=6(x_2+2), &\\
	&\qquad\sigma_{10}=6(x_3+2), &\\
	&\qquad\sigma_{11}=6, & s^6\rightarrow \lambda, &\\
\\
&t_{j+5}:\; \\
	&\qquad\sigma_{2},\sigma_{3},\sigma_{4}=21(h+i)-8,	&s^{21(h+i)-8}/s^3\rightarrow s,\\
	&\qquad\sigma_{5},\sigma_{6},\sigma_{7}=21l-16,	&\\
	&\qquad\sigma_{8}=6(x_1+4), &\\
	&\qquad\sigma_{9}=6(x_2+3), &\\
	&\qquad\sigma_{10}=6(x_3+3),&\\
	&\qquad\sigma_{11}=6, & s^6\rightarrow \lambda.
\end{xalignat*}
In configurations $t_{j+3}$, $t_{j+4}$ and $t_{j+5}$ each of the simulated counters $\sigma_{8}$, $\sigma_{9}$ and $\sigma_{10}$ are incremented. In configurations $t_{j+6}$ to $t_{j+10}$ the simulated counter $\sigma_{8}$ is decremented 4 times and the simulated counters $\sigma_{9}$ and $\sigma_{10}$ are each decremented twice. 
\begin{xalignat*}{2}
&t_{j+6}:\; \\
	&\qquad\sigma_{2},\sigma_{3}=21(h+i)-10,	&s^{21(h+i)-10}/s^5\rightarrow s,\\
	&\qquad\sigma_{4}=21(h+i)-10,	&s^{21(h+i)-10}/s^{21(h+i-l)+5}\rightarrow s,\\
	&\qquad\sigma_{5},\sigma_{6},\sigma_{7}=21l-15,	&\\
	&\qquad\sigma_{8}=6(x_1+4)+3, &(s^6)^{\ast}s^{9}/s^{9}\rightarrow s,\\
	&\qquad\sigma_{9}=6(x_2+3)+3, &(s^6)^{\ast}s^{9}/s^{9}\rightarrow s,\\
	&\qquad\sigma_{10}=6(x_3+3)+3, &(s^6)^{\ast}s^{15}/s^3\rightarrow s,\\
	&\qquad\sigma_{11}=3, & s^3\rightarrow \lambda. &
\end{xalignat*}
\begin{xalignat*}{2}
&t_{j+7}:\; \\
	&\qquad\sigma_{2},\sigma_{3}=21(h+i)-11,	&s^{21(h+i)-11}/s^6\rightarrow s,\\
	&\qquad\sigma_{4},\sigma_{5},\sigma_{6},\sigma_{7}=21l-11,	&\\
	&\qquad\sigma_{8}=6(x_1+3)+3, &(s^6)^{\ast}s^{9}/s^{9}\rightarrow s,\\
	&\qquad\sigma_{9}=6(x_2+2)+3, &(s^6)^{\ast}s^{9}/s^{9}\rightarrow s,\\
	&\qquad\sigma_{10}=6(x_3+3)+3, &(s^6)^{\ast}s^{15}/s^3\rightarrow s,\\
	&\qquad\sigma_{11}=4, & s^4\rightarrow \lambda, &\\
\\
&t_{j+8}:\; \\
	&\qquad\sigma_{2},\sigma_{3}=21(h+i)-13,	&s^{21(h+i)-13}/s^{21(h+i-l)-6}\rightarrow s,\\
	&\qquad\sigma_{4},\sigma_{5},\sigma_{6},\sigma_{7}=21l-7,	&\\
	&\qquad\sigma_{8}=6(x_1+2)+2, &(s^6)^{\ast}s^{8}/s^{8}\rightarrow s,\\
	&\qquad\sigma_{9}=6(x_2+1)+2, &(s^6)^{\ast}s^{14}/s^{2}\rightarrow s,\\
	&\qquad\sigma_{10}=6(x_3+3)+2, &(s^6)^{\ast}s^{8}/s^{8}\rightarrow s,\\
	&\qquad\sigma_{11}=3, & s^3\rightarrow \lambda, &\\
\\
&t_{j+9}:\; \\
	&\qquad\sigma_{2},\sigma_{3},\sigma_{4},\sigma_{5},\sigma_{6},\sigma_{7}=21l-3,	&\\
	&\qquad\sigma_{8}=6(x_1+1)+2, &(s^6)^{\ast}s^{8}/s^{8}\rightarrow s,\\
	&\qquad\sigma_{9}=6(x_2+1)+2, &(s^6)^{\ast}s^{14}/s^{2}\rightarrow s,\\
	&\qquad\sigma_{10}=6(x_3+2)+2, &(s^6)^{\ast}s^{8}/s^{8}\rightarrow s,\\
	&\qquad\sigma_{11}=3, & s^3\rightarrow \lambda, &\\
\\
&t_{j+10}:\; \\
	&\qquad\sigma_{2},\sigma_{3},\sigma_{4},\sigma_{5},\sigma_{6},\sigma_{7}=21l,	&\\
	&\qquad\sigma_{8}=6x_1, &\\
	&\qquad\sigma_{9}=6(x_2+1), &\\
	&\qquad\sigma_{10}=6(x_3+1), &\\
	&\qquad\sigma_{11}=1, & s\rightarrow \lambda, &\\
	&\qquad\sigma_{12},\sigma_{13},\sigma_{14}=21h-29, &(s^{3})^{\ast}s^4/s^3\rightarrow s,\\ 
	&\qquad\sigma_{15},\sigma_{16},\sigma_{17}=30. &
\end{xalignat*}
Note that at time $t_{j+8}$ that rule $(s^6)^{\ast}s^{14}/s^{2}\rightarrow s$ will always be applicable as here $x_2>0$ (see the second line at the start of the proof). 
\newpage
The remainder of this simulation is similar to the computation carried out at the end of the initialisation process (see the last paragraph of Section~\ref{sec:Algorithm overview Pi'_(C_3)} and timesteps $t_{x_1+x_2+4}$ to $t_{x_1+x_2+14h+3}$ of the Section~\ref{sec:Encoding of a configuration of C_3 and reading in input to Pi'_(C_3)}). Thus, after a further $14h-8$ timesteps we get
\begin{xalignat*}{2}
&t_{j+14h+2}:\; \\
	&\qquad\sigma_{2},\sigma_{3},\sigma_{4},\sigma_{5},\sigma_{6},\sigma_{7}=21(h+l)+1,	&\\
	&\qquad\sigma_{8}=6x_1, &\\
	&\qquad\sigma_{9}=6(x_2+1), &\\
	&\qquad\sigma_{10}=6(x_3+1), &\\	&\qquad\sigma_{12},\sigma_{13},\sigma_{14}=21h+1, &(s^{3})^{\ast}s^4/s^3\rightarrow s.
\end{xalignat*}
At timestep $t_{j+14h+2}$ the simulation of $q_i:DEC(1),q_l,q_k$ for Case 1 ($x_1>0$) is complete.  Note that a decrement on the value $x_1$ in counter $c_1$ is simulated by decreasing the value in $\sigma_{8}$ from $6(x_1+1)$ to $6x_1$. Note also that the encoding $21(h+l)+1$ of the next instruction $q_{l}$ has been established in neurons $\sigma_{2}$, $\sigma_{3}$, $\sigma_{4}$, $\sigma_{5}$, $\sigma_{6}$ and $\sigma_{7}$. Alternatively, if we have Case 2 ($x_1=0$) then we get 
\begin{xalignat*}{2}
&t_{j+1}:\; \\
	&\qquad\sigma_{2}=21(h+i)-4,	&\\
	&\qquad\sigma_{3},\sigma_{4},\sigma_{5},\sigma_{6},\sigma_{7}=21(h+i)+2, &\\
	&\qquad\sigma_{8}=7, &s^7\rightarrow \lambda,\\
	&\qquad\sigma_{9}=6(x_2+1)+1, &(s^6)^{\ast}s^7/s^7\rightarrow s,\\
	&\qquad\sigma_{10}=6(x_3+1)+1, &(s^6)^{\ast}s^7/s^7\rightarrow s,\\ 
	&\qquad\sigma_{11}=1, & s\rightarrow \lambda, &\\
\\
&t_{j+2}:\; \\
	&\qquad\sigma_{2}=21(h+i)-2,	&s^{21(h+i)-2}/s^{21(h+i-k)-1}\rightarrow s,\\
	&\qquad\sigma_{3},\sigma_{4},\sigma_{5},\sigma_{6},\sigma_{7}=21(h+i)+4, &s^{21(h+i)+4}/s^{21(h+i-k)+5}\rightarrow s,\\
	&\qquad\sigma_{9}=6x_2, &\\
	&\qquad\sigma_{10}=6x_3,\\
	&\qquad\sigma_{11}=1, & s\rightarrow \lambda, &\\
\\
&t_{j+3}:\; \\
	&\qquad\sigma_{2},\sigma_{3},\sigma_{4},\sigma_{5},\sigma_{6},\sigma_{7}=21k, &\\
	&\qquad\sigma_{8}=6, &\\
	&\qquad\sigma_{9}=6(x_2+1), &\\
	&\qquad\sigma_{10}=6(x_3+1),\\
	&\qquad\sigma_{11}=6, & s^6\rightarrow \lambda, &\\
	&\qquad\sigma_{12},\sigma_{13},\sigma_{14}=21h-8, &(s^{3})^{\ast}s^4/s^3\rightarrow s,\\ 
	&\qquad\sigma_{15},\sigma_{16},\sigma_{17}=9. &\\ 
\end{xalignat*}
The remainder of this simulation is similar to the computation carried out at the end of the initialisation process (see the last paragraph of Section~\ref{sec:Algorithm overview Pi'_(C_3)} and timesteps $t_{x_1+x_2+4}$ to $t_{x_1+x_2+14h+3}$ of the Section~\ref{sec:Encoding of a configuration of C_3 and reading in input to Pi'_(C_3)}). Thus, after a further $14h-1$ timesteps we get
\begin{xalignat*}{2}
&t_{j+14h+2}:\; \\
	&\qquad\sigma_{2},\sigma_{3},\sigma_{4},\sigma_{5},\sigma_{6},\sigma_{7}=21(h+k)+1, &\\
	&\qquad\sigma_{9}=6, &\\
	&\qquad\sigma_{9}=6(x_2+1), &\\
	&\qquad\sigma_{10}=6(x_3+1),\\
	&\qquad\sigma_{12},\sigma_{13},\sigma_{14}=21h+1.\\ 
\end{xalignat*}
At time $t_{j+14h+2}$ the simulation of $q_i:DEC(1),q_l,q_k$ for Case 2 ($x_1=0$), is complete. Note that the encoding $21(h+k)+1$ of the next instruction $q_{k}$ has been established in neurons $\sigma_{2}$, $\sigma_{3}$, $\sigma_{4}$, $\sigma_{5}$, $\sigma_{6}$ and $\sigma_{7}$.

\subsubsection{Halting}
If $C_3$ enters the halt instruction $q_h$ at time $t_j$ then we get the following
\begin{xalignat*}{2}
&t_{j}:\; \\
	&\qquad\sigma_{2},\sigma_{3},\sigma_{4},\sigma_{5}=42h+1,	&s^{42h+1}/s\rightarrow s,\\
	&\qquad\sigma_{6},\sigma_{7}=42h+1,	&\\
	&\qquad\sigma_{8}=6(x_1+1), &\\
	&\qquad\sigma_{9}=6(x_2+1), &\\
	&\qquad\sigma_{10}=6(x_o+1), &\\
\\
&t_{j+1}:\; \\
	&\qquad\sigma_{2},\sigma_{3},\sigma_{4},\sigma_{5}=42h+1,	&s^{42h+1}/s\rightarrow s,\\
	&\qquad\sigma_{6},\sigma_{7}=42h+2,	&\\
	&\qquad\sigma_{8}=6(x_1+1)+4, &\\
	&\qquad\sigma_{9}=6(x_2+1)+4, &(s^{6})^{\ast}s^{10}/s^{6}\rightarrow s,\\
	&\qquad\sigma_{10}=6(x_o+1)+4, &(s^{6})^{\ast}s^{10}/s^{6}\rightarrow s,\\
	&\qquad\sigma_{11}=4, &s^4\rightarrow \lambda,\\
\\
&t_{j+2}:\; \\
	&\qquad\sigma_{2},\sigma_{3}=42h+3,	&s^{\ast}s^{42h+3}/s\rightarrow s,\\
	&\qquad\sigma_{4},\sigma_{5}=42h+3,	&\\
	&\qquad\sigma_{6},\sigma_{7}=42h+5,	&\\
	&\qquad\sigma_{8}=6(x_1+2)+2, &(s^{6})^{\ast}s^{8}/s^{8}\rightarrow s,\\
	&\qquad\sigma_{9}=6(x_2+1)+2, &(s^{6})^{\ast}s^{14}/s^{2}\rightarrow s,\\
	&\qquad\sigma_{10}=6(x_o+1)+2, &(s^{6})^{\ast}s^{8}/s^{8}\rightarrow s,\\
	&\qquad\sigma_{11}=5. &
\end{xalignat*}
Note that after time $t_{j+2}$ we can ignore neurons $\sigma_{4}$, $\sigma_{5}$, $\sigma_{6}$ and $\sigma_{7}$ as there are no rules applicable in these neurons when the number of spikes is $\geqslant 43h+3$. The number of spikes in $\sigma_{2}$ and $\sigma_{3}$ does not decrease following timestep $t_{j+2}$, and thus the rule $s^{\ast}s^{42h+3}/s\rightarrow s$ is applicable at each subsequent timestep regardless of the operation of neurons $\sigma_{8}$ and $\sigma_{9}$. Thus, neurons $\sigma_{8}$ and $\sigma_{9}$ may also be ignored as their operation has no effect on the remainder of the simulation. Note that in subsequent configurations we write $\sigma_{2},\sigma_{3}\geqslant 42h+3$ as there are more than $42h+3$ spikes in each of these neurons. Thus we have
\begin{xalignat*}{2}
&t_{j+3}:\; \\
	&\qquad\sigma_{2},\sigma_{3}\geqslant 42h+3,	&s^{\ast}s^{42h+3}/s\rightarrow s,\\
	&\qquad\sigma_{10}=6x_o+2, &(s^{6})^{\ast}s^{8}/s^{8}\rightarrow s,\\
	&\qquad\sigma_{11}=8, &\\
\\
&t_{j+4}:\; \\
	&\qquad\sigma_{2},\sigma_{3}\geqslant 42h+3,	&s^{\ast}s^{42h+3}/s\rightarrow s,\\
	&\qquad\sigma_{10}=6(x_o-1)+2, &(s^{6})^{\ast}s^{8}/s^{8}\rightarrow s,\\
	&\qquad\sigma_{11}=11, &s^{11}/s^2\rightarrow s,\\
\\
&t_{j+5}:\; \\
	&\qquad\sigma_{2},\sigma_{3}\geqslant 42h+3,	&s^{\ast}s^{42h+3}/s\rightarrow s,\\
	&\qquad\sigma_{10}=6(x_o-2)+2, &(s^{6})^{\ast}s^{8}/s^{8}\rightarrow s,\\
	&\qquad\sigma_{11}=12. &
\end{xalignat*}
The rule $(s^{6})^{\ast}s^{8}/s^{8}\rightarrow s$ is applied in $\sigma_{10}$ a further $x_o-2$ times until we get
\begin{xalignat*}{2}
&t_{j+x_o+3}:\; \\
	&\qquad\sigma_{2},\sigma_{3}\geqslant 42h+3,	&s^{\ast}s^{42h+3}/s\rightarrow s,\\
	&\qquad\sigma_{10}=2, &s^{2}\rightarrow \lambda,\\
	&\qquad\sigma_{11}=3(x_o-2)+12, &\\
\\
&t_{j+x_o+4}:\; \\
	&\qquad\sigma_{2},\sigma_{3}\geqslant 42h+3,	&s^{\ast}s^{42h+3}/s\rightarrow s,\\
	&\qquad\sigma_{10}=2, &s^{2}\rightarrow \lambda,\\
	&\qquad\sigma_{11}=3(x_o-2)+14, &(s^3)^{\ast}s^{14}/s\rightarrow s.
\end{xalignat*}
Recall from Section~\ref{sec:Spiking neural P-systems} that the output of an SN P system is the time interval between the first and second spikes that are sent out of the output neuron. Note from above that the output neuron $\sigma_{11}$ fires for the first time at timestep $t_{j+4}$ and for the second time at timestep $t_{j+x_o+4}$. Thus, the output of $\Pi'_{C_3}$ is $x_o$ the contents of the output counter $c_3$ when $C_3$ enters the halt instruction $q_h$. If $x_o=0$ neuron $\sigma_{11}$ will fire only once. To see this, note that if $x_o=0$ then $s^{2}\rightarrow \lambda$ will be applied in neuron $\sigma_{10}$ at time $t_{j+3}$, and thus $\sigma_{11}$ will have 10 spikes (instead of 11) at time $t_{j+4}$ and the rule $s^{10}\rightarrow s$ will be applied in $\sigma_{11}$ ending the computation.

\begin{table}[t]
\begin{center}
\begin{tabular}{l@{\;\;\;}|@{\;\;\;}l}
origin neurons	& target neurons \\ \hline 
$\sigma_1$ & $\sigma_2,\sigma_3,\sigma_4,\sigma_5,\sigma_6,\sigma_7,\sigma_8,\sigma_9,\sigma_{10},\sigma_{12},\sigma_{13},\sigma_{14}$	\\
$\sigma_2,$ & $\sigma_3,\sigma_4,\sigma_5,\sigma_6,\sigma_7,\sigma_8,\sigma_9,\sigma_{10},\sigma_{11}$\\
$\sigma_3,$ & $\sigma_2,\sigma_8,\sigma_9,\sigma_{10},\sigma_{11}$\\
$\sigma_4,\sigma_5,\sigma_6,\sigma_7$ & $\sigma_8,\sigma_9,\sigma_{10},\sigma_{11}$\\
$\sigma_8,\sigma_9$ & $\sigma_2,\sigma_3,\sigma_4,\sigma_5,\sigma_6,\sigma_7$	\\
$\sigma_{10}$ & $\sigma_2,\sigma_3,\sigma_4,\sigma_5,\sigma_6,\sigma_7,\sigma_{11}$	\\
$\sigma_{12},\sigma_{13},\sigma_{14}$ & $\sigma_{15},\sigma_{16},\sigma_{17}$	\\
$\sigma_{15},\sigma_{16},\sigma_{17}$ & $\sigma_2,\sigma_3,\sigma_4,\sigma_5,\sigma_6,\sigma_7,\sigma_{12},\sigma_{13},\sigma_{14}$	\\\hline 
\end{tabular}
\end{center}
\caption{This table gives the set of synapses of the SN P system $\Pi'_{C_3}$. Each origin neuron $\sigma_i$ and target neuron $\sigma_j$ that appear on the same row have a synapse going from $\sigma_i$ to $\sigma_j$.}\label{tab:synapses}
\end{table}

We have shown how to simulate arbitrary instructions of the form $q_i:INC(1),q_l$ and $q_i:DEC(1),q_l,q_k$. Instructions that operate on counters $c_2$ and $c_3$ are simulated in a similar manner. Immediately following the simulation of an instruction $\Pi'_{C_3}$ is configured to begin simulation of the next instruction. Each instruction of $C_3$ is simulated in $14h+2$ timesteps.  The pair of input values ($x_1,x_2$) is read into the system in $x_1+x_2+14h+3$ timesteps and sending the output value $x_o$ out of the system takes $x_o+4$ timesteps. Thus, if $C_3$ completes it computation in time $t$ then $\Pi'_{C_3}$ simulates the computation of $C_3$ in linear time $O(ht+x_1+x_2+x_o)$. \qed
\end{pf}

\section{Lower bounds for small universal SN P systems}\label{sec:Lower bounds}
In this section we show that there exists no universal SN P system with only 3 neurons even when we allow the input technique to be generalised. This is achieved in Theorem~\ref{thm:no universal SNP system with 3 neurons} by showing that these systems are simulated by log-space bounded Turing machines. Following this, we show that if we generalise the output technique we can give a universal SN P system with extended rules that has only 3 neurons. As a corollary of our proof of Theorem~\ref{thm:no universal SNP system with 3 neurons}, we find that a universal SN P system with extended rules and generalised input and output is not possible with 2 neurons.

          \begin{figure}[t]
                    \begin{tikzpicture}
			\tikzstyle{state}=[ellipse,draw=black!75, node distance=1.5cm]
			\tikzstyle{state1}=[ellipse,draw=black!75, node distance=1.8cm]
                        \tikzstyle{dots}=[draw=none,node distance=1cm]
                        \tikzstyle{dots1}=[draw=none,node distance=1.2cm]

                        \node[state] (g1)         		{\small $g_1$};
                        \node[state] (g2)    	[right of=g1]  	{\small $g_2$};
                        \node[state] (g3)    	[right of=g2]	{\small $g_3$};
                        \node[dots]  (dots1) 	[right of=g3] 	{\ldots};
                        \node[state] (gu-1)  	[right of=dots1,node distance=1cm] {\small $g_{u-1}$};
                        \node[state1](gu)	[right of=gu-1] 	{\small $g_u$};
                        \node[state1](gu+1) 	[right of=gu] 	{\small $g_{u+1}$};
                        \node[dots1]  (dots2) 	[right of=gu+1]	{\ldots};
                        \node[state] (gv)    	[right of=dots2,node distance=1cm] {\small $g_v$};

                        \path [->] (g1)   edge              node  {} (g2)
                                   (g2)   edge              node  {} (g3)
                                   (gu-1) edge              node  {} (gu)
                                   (gu)   edge              node  {} (gu+1)
                                   (gv)   edge [bend right,in=270] node [above] {$s$} (gu);

			\draw (g1)+(.75,.15) node {$s$};
			\draw (g2)+(.75,.15) node {$s$};
			\draw (gu-1)+(1,.15) node {$s$};
			\draw (gu)+(.75,.15) node {$s$};
                        \draw (g3)+(.7,1.1) node {\Large $G$};
                    \end{tikzpicture}
     \begin{tikzpicture}
			\tikzstyle{state}=[ellipse,draw=black!75, node distance=1.5cm]
			\tikzstyle{state1}=[ellipse,draw=black!75, node distance=1.8cm]
                        \tikzstyle{dots}=[draw=none,node distance=1cm]
                        \tikzstyle{dots1}=[draw=none,node distance=1.2cm]

                        \node[state] (g1)         		{\small $g_1$};
                        \node[state] (g2)    	[right of=g1]  	{\small $g_2$};
                        \node[state] (g3)    	[right of=g2]	{\small $g_3$};
                        \node[dots]  (dots1) 	[right of=g3] 	{\ldots};
                        \node[state] (gu-1)  	[right of=dots1,node distance=1cm] {\small $g_{u-1}$};
                        \node[state1](gu)	[right of=gu-1] 	{\small $g_u$};
                        \node[state1](gu+1) 	[right of=gu] 	{\small $g_{u+1}$};
                        \node[dots1]  (dots2) 	[right of=gu+1]	{\ldots};
                        \node[state] (gv)    	[right of=dots2,node distance=1cm] {\small $g_v$};

                        \path [->] 	(g1)   	edge [bend left,in=150] node  [above] {$+s$} (g2)
                                   	(g2)   	edge [bend left,in=150] node  [below] {$-s$} (g1)
                                   	(g2)   	edge [bend left,in=150] node  [above] {$+s$} (g3)
                                   	(g3)   	edge [bend left,in=150] node  [below] {$-s$} (g2)
                                   	(gu-1) 	edge [bend left,in=150] node  [above] {$+s$} (gu)
                                   	(gu) 	edge [bend left,in=150] node  [below] {$-s$} (gu-1)
                                   	(gu)   	edge [bend left,in=140] node  [above] {$+s$} (gu+1)
                                   	(gu+1)	edge [bend left,in=164] node  [above] {$-s$} (gu)
                    			(gv)   	edge [bend right,in=270] node [above] {$+s$} (gu)
                    			(gu)   	edge [bend right,in=270] node [below] {$-s$} (gv);
                        \draw (g3)+(.7,1.1) node {\Large $G'$};
                    \end{tikzpicture}
            \caption{Finite state machine $G$ decides if there is any rule applicable in a neuron given the number of spikes in the neuron \emph{at a given time} in the computation. Each $s$ represents a spike in the neuron. Machine $G'$ keeps track of the movement of spikes into and out of the neuron and decides whither or not a particular rule is applicable \emph{at each timestep} in the computation. $+s$ represents a single spike entering the neuron and $-s$ represents a single spike exiting the neuron.}\label{fig:finite_state_machine_G}
\end{figure}
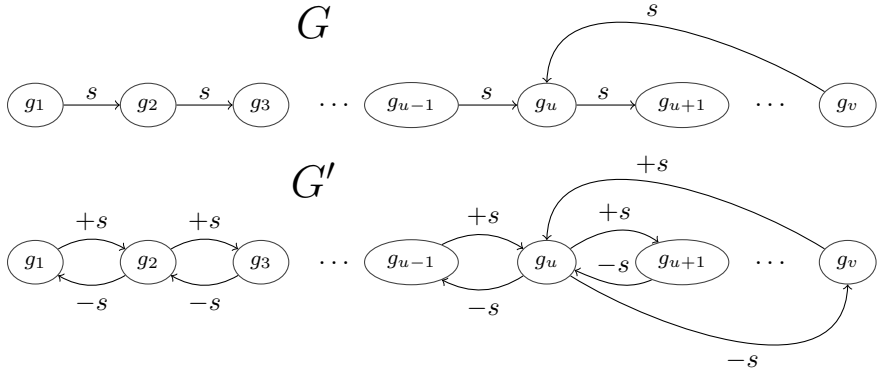

In this and other work~\cite{Paun2007,Zhang2008B} on small SN P systems the input neuron only receives a constant number of spikes from the environment and the output neuron fires no more than a constant number of times. Hence, we call input standard if the input neuron receives no more than $y$ spikes from the environment, where $y$ is a constant independent of the input (i.e. the number of 1s in its input sequence is $<y$). Similarly, we call the output standard if the output neuron fires no more than $x$ times, where $x$ is a constant independent of the input. Here we say an SN P system has generalised input if the input neuron is permitted to receive $\leqslant n$ spikes from the environment where $n\in\Nset$ is the length of its input sequence.
 
\begin{theorem}\vspace{.3cm}\label{thm:no universal SNP system with 3 neurons}
Let $\Pi$ be any extended SN P system with only 3 neurons, generalised input and standard output. Then there is a non-deterministic Turing machine $T_{\Pi}$ that simulates the computation of $\Pi$ in space $O(\log n)$ where $n$ is the length of the input to $\Pi$.
\end{theorem}
\begin{pf}
Let $\Pi$ be any extended SN P system with generalised input, standard output, and neurons $\sigma_1$, $\sigma_2$ and $\sigma_3$. Also, let $x$ be the maximum number of times the output neuron $\sigma_3$ is permitted to fire and let $q$ and $r$ be the maximum value for $b$ and $p$ respectively, for all $E/s^b\rightarrow s^p;d$ in $\Pi$.

We begin by explaining how the activity of $\sigma_3$ may be simulated using only the states of $T_{\Pi}$ (i.e. no workspace is required to simulate $\sigma_3$). Recall that the applicability of each rule is determined by a regular expression over a unary alphabet. We can give a single regular expression $R$ that is the union of all the regular expressions for the firing rules of $\sigma_3$. This regular expression $R$ determines whither or not there is any applicable rule in $\sigma_3$ at each timestep. Figure~\ref{fig:finite_state_machine_G} gives the deterministic finite automata $G$ that accepts $L(R)$ the language generated by $R$. During a computation we may use $G$ to decide which rules are applicable in $\sigma_i$ by passing an $s$ to $G$ each time a spike enters $\sigma_3$. However, $G$ may not give the correct result if spikes leave the neuron as it does not record spikes leaving $\sigma_i$. Thus, using $G$ we may construct a second machine $G'$ such that $G'$ records the movement of spikes going into and out of the neuron. $G'$ is construct as follows: $G'$ has all the same states (including accept states) and transitions as $G$ along with an extra set of transitions that record spikes leaving the neuron. This extra set of transitions are given as follows: for each transition on $s$ from a state $g_i$ to a state $g_j$ in $G$ there is a new transition on $-s$ going from state $g_j$ to $g_i$ in $G'$ that records the removal of a spike from $\sigma_3$. By recording the dynamic movement of spikes, $G'$ is able to decide which rules are applicable in $\sigma_3$ at each timestep during the computation. $G'$ is also given in Figure~\ref{fig:finite_state_machine_G}. To simulate the operation of $\sigma_3$ we emulate the operation of $G'$ in the states of $T_{\Pi}$. Note that there is a single non-deterministic choice to be made in $G'$. This choice is at state $g_u$ if a spike is being removed ($-s$). It would seem that in order to make the correct choice in this situation we need to know the exact number of spikes in $\sigma_3$. However, we need only store at most $u+yq$ spikes. The reason for this is that if there are $\geqslant u+yq$ spikes in $\sigma_3$, then $G'$ will not enter state $g_{u-1}$ again. To see this, note that $\sigma_3$ spikes a maximum of $y$ times using at most $q$ spikes each time, and so once there are $>u+yq$ spikes the number of spikes in $\sigma_3$ will be $>u-1$ for the remainder of the computation. Thus, $T_{\Pi}$ simulates the activity of $\sigma_3$ by simulating the operation of $G'$ and encoding at most $u+yq$ spikes in its states.

In this paragraph we explain the operation of $T_{\Pi}$. Following this, we give an analysis of the space complexity of $T_{\Pi}$. $T_{\Pi}$ has 4 tapes including an output tape, which is initially blank, and a read only input tape. 
The tape head on both the input and output tapes is permitted to only move right. Each of the remaining tapes, tapes 1 and 2 simulate the activity of the neurons $\sigma_1$ and $\sigma_2$, respectively. 
These tapes record the number of spikes in $\sigma_1$ and $\sigma_2$. A timestep of $\Pi$ is simulated as follows: $T_{\Pi}$ scans tapes 1 and 2 to determine if there are any applicable rules in $\sigma_1$ and $\sigma_2$ at that timestep. 
The applicability of each neural rule in $\Pi$ is determined by a regular expression and so a decider for each rule is easily implemented in the states of $T_{\Pi}$. 
Recall from the previous paragraph that the applicability of the rules in $\sigma_3$ is already recorded in the states of $T_{\Pi}$. 
Also, $T_{\Pi}$ is non-deterministic and so if more than one rule is applicable in a neuron $T_{\Pi}$ simply chooses the rule to simulate in the same manner as $\Pi$. 
Once $T_{\Pi}$ has determined which rules are applicable in each of the three neurons at that timestep it changes the encodings on tapes 1 and 2 to simulate the change in the number of spikes in neurons $\sigma_1$ and $\sigma_2$ during that timestep. 
As mentioned in the previous paragraph any change in the number of spikes in $\sigma_3$ is recorded in the states of $T_{\Pi}$. 
The input sequence of $\Pi$ may be given as binary input to $T_{\Pi}$ by placing it on its input tape. Also, if at a given timestep a 1 is read on the input tape then $T_{\Pi}$ simulates a spike entering the simulated input neuron. At each simulated timestep, if the output neuron $\sigma_3$ spikes then a 1 is place on the output tape, and if $\sigma_3$ does not spike a 0 is placed on the output tape. Thus the output of $\Pi$ is encoded on the output tape when the simulation ends.

In a two neuron system each neuron has at most one out-going synapse and so the number of spikes in the system does not increase over time. Thus, the total number of spikes in neurons $\sigma_1$ and $\sigma_2$ can only increase when $\sigma_3$ fires or a spike is sent into the system from the environment. The input is of length $n$, and so $\sigma_1$ and $\sigma_2$ receive a maximum of $n$ spikes from the environment. Neuron $\sigma_3$ fires a total of $y$ times sending at most $r$ spikes each time and so the maximum number of spikes in $\sigma_1$ and $\sigma_2$ during the computation is $n+2ry$. Using a binary encoding tapes 1 and 2 of $T_{\Pi}$ encode the number of spikes in $\sigma_1$ and $\sigma_2$ using space of $\log_2 (n+2ry)$. As mentioned earlier no space is used to simulate $\sigma_3$, and thus $T_{\Pi}$ simulates $\Pi$ using space of $O(\log n)$. \qed
\end{pf}

It is interesting to note that with a slight generalisation on the system in Theorem~\ref{thm:no universal SNP system with 3 neurons} we obtain universality. If we remove the restriction that allows the output neuron to fire only a constant number of times then we may construct a universal SN P system with extended rules and only three neurons. Here we define the output of an extended SN P system with generalised output to the time interval between the first and second timesteps where exactly $x$ spikes are sent out of the output neuron. \vspace{.3cm}
\begin{theorem}\label{thm:universal Extended SNP system with generalised output and 3 neuron}
Let $C_2$ be a universal counter machine with 2 counters that completes it computation in time $t$ to give the output value $x_o$ when given the input value $x_1$. Then there is a universal extended SN P system $\Pi''_{C_2}$ with standard input and generalised output that simulates the computation of $C_2$ in time $O(t+x_1+x_o)$ and has only 3 neurons. 
\end{theorem}
\begin{pf}
A graph of $\Pi''_{C_2}$ is constructed by removing the output neuron $\sigma_5$ from the system $\Pi_{C_2}$ given in the proof of Theorem~\ref{thm:universal Extended SNP system with 4 neuron} and making $\sigma_3$ the new output neuron of $\Pi''_{C_2}$. The rules for $\Pi''_{C_2}$ are given by the first 3 rows of Table~\ref{tab:neurons of Extended SNP 2-counter} and a diagram of the system is obtained by removing neurons $\sigma_4$ and $\sigma_5$ from Figure~\ref{fig:extended universal SNP system} and adding a synapse to the environment from the new output neuron $\sigma_3$. The operation of $\Pi''_{C_2}$ is identical to the operation of $\Pi_{C_2}$ with the exception of the new output technique. The output of $\Pi''_{C_2}$ is the time interval between the first and second timesteps where exactly $2$ spikes are sent out of the output neuron~$\sigma_3$. \qed
\end{pf}
From the third paragraph of the proof of Theorem~\ref{thm:no universal SNP system with 3 neurons} we get the following immediate corollary.\vspace{.3cm}
\begin{corollary}\label{cor:no universal extended SNP system with generalised output and 2 neuron}
Let $\Pi$ be any extended SN P system with only 2 neurons and generalised input and output. Then there is a non-deterministic Turing machine $T_{\Pi}$ that simulates the computation of $\Pi$ in space $O(\log n)$ where $n$ is the length of the input to $\Pi$.
\end{corollary}

\section{Conclusion}
The dramatic improvement on the size of earlier small universal SN P system given by Theorems~\ref{thm:universal extended SNP systems with 5 neurons} and~\ref{thm:universal SNP systems with 17 neuron} is in part due to the method we use to encode the instructions of the counter machines our systems simulate. In the systems of P{\u{a}}un and P{\u{a}}un~\cite{Paun2007} each counter machine instruction was encoded by a unique set of neurons. Thus the size of the system is dependant on the number of instructions in the counter machine being simulated. Some improvement was made by Zhang et al.~\cite{Zhang2008B} by showing that certain types of instructions may be grouped together. However, the number of neurons used by the system remained dependant on the number of instructions in the counter machine being simulated. In our systems each unique counter machine instruction is encoded as a unique number of spikes and thus the size of our SN P systems are independent of the number of instruction used by the counter machine they are simulating. The technique of encoding the instructions as spikes was first used to construct small universal SN P systems in~\cite{Neary2008B}.

The results from Theorems~\ref{thm:universal Extended SNP system with 4 neuron} and~\ref{thm:no universal SNP system with 3 neurons} give tight upper and lower bounds on the size of the smallest universal SN P system with extended rules. Thus in Theorem~\ref{thm:universal Extended SNP system with 4 neuron} we have given the smallest possible universal SN P system with extended rules. The results from Theorem~\ref{thm:universal Extended SNP system with generalised output and 3 neuron} and Corollary~\ref{cor:no universal extended SNP system with generalised output and 2 neuron} give tight upper and lower bounds on the size of the smallest universal SN P systems with extended rules and generalised output. Thus, Theorem~\ref{thm:universal Extended SNP system with generalised output and 3 neuron} gives the smallest possible universal SN P system with extended rules and generalised output. 

The lower bounds given in Theorem~\ref{thm:no universal SNP system with 3 neurons} are also applicable to standard SN P systems and thus give a lower bound of 4 neurons for the smallest possible standard system that is universal. However, when compared with extended systems the rules used in standard SN P systems are quite limited, and so it seems likely that this lower bound of 4 neurons can be increased.  Note that here and in~\cite{Paun2007,Zhang2008B} the size of a universal SN P system is measured by the number of neurons in the system. However, the size of an SN P system could also be measured by the number of neural rules in the system.

\bibliographystyle{abbrv} 

\bibliography{2010_Neary_arXiv_SNP1}

\begin{table}[h]
\begin{center}
\begin{tabular}{c@{\;\;\;}|@{\;\;\;}l}
neuron       	& rules  \\ \hline
$\sigma_1$   	& $s^{8h+1}/s^{8h}\rightarrow s^{8h},$\qquad   $s^{8h+2}/s^{8h+1}\rightarrow s^{8h+1},$ \qquad   $s^{6h+1}\rightarrow s^{4h+4},$\qquad $s^{2}\rightarrow \lambda,$\qquad   $s\rightarrow \lambda$ \\ 
		& $s^{16h+4i}\rightarrow s^{12h+4l},$\qquad  $s^{10h+4i}\rightarrow s^{4(h+l)},$\qquad if $l<h$ \\ 
		& $s^{16h+4i}\rightarrow s^{12h+5},$\qquad  $s^{10h+4i}\rightarrow s^{4h+5},$\qquad if $l=h$ \\
& $s^{8h+4i}\rightarrow s^{4(h+k)},$\qquad  if $k\neq h$ \\
& $s^{8h+4i}\rightarrow s^{4h+5},$\qquad  if $k=h$ \\ \hline
$\sigma_2$	& $(s^{8h})^{\ast}s^{8h+1}/s^{8h}\rightarrow s,$\qquad   $(s^{8h})^{\ast}s^{8h+2}/s^{8h+2}\rightarrow s^{2h}$ \\ 
		& $(s^{8h})^{\ast}s^{4(h+i)}/s^{4(h+i)}\rightarrow s^{4(h+i)}$\quad if $q_i:INC(1)\in\{Q\}$ \\ 
		& $(s^{8h})^{\ast}s^{4(h+i)}/s^{8h+4(h+i)}\rightarrow s^{6h}$\quad if $q_i:INC(x)\in\{Q\},\, x\neq 1$ \\
		& $(s^{8h})^{\ast}s^{16h+4(h+i)}/s^{12h+4i}\rightarrow s^{6h+4i}$\quad if $q_i:DEC(1)\in\{Q\}$ \\ 
		& $s^{8h+4(h+i)}/s^{4(h+i)}\rightarrow s^{4(h+i)}$\quad if $q_i:DEC(1)\in\{Q\}$ \\  
		& $(s^{8h})^{\ast}s^{4(h+i)}/s^{4(h+i)}\rightarrow s^{2h}$\quad if $q_i:DEC(x)\in\{Q\},\, x\neq 1$ \\ \hline
$\sigma_3$	& $s^{2}/s\rightarrow s,$\qquad   $s^{8h+1}/s^{8h}\rightarrow s,$ \qquad   $(s^{8h})^{\ast}s^{8h+3}/s^{8h+3}\rightarrow s^{2h},$ \qquad $(s^{8h})^{\ast}s^{20h+5}/s^{12h}\rightarrow s^{2}$\\
		&  $(s^{8h})^{\ast}s^{16h+5}/s^{8h}\rightarrow s,$ \qquad $s^{8h+5}\rightarrow s^2,$\qquad $s^{12h+5}\rightarrow s^2$ \\
		& $(s^{8h})^{\ast}s^{4(h+i)}/s^{4(h+i)}\rightarrow s^{4(h+i)}$\quad if $q_i:INC(2)\in\{Q\}$ \\ 
		& $(s^{8h})^{\ast}s^{4(h+i)}/s^{8h+4(h+i)}\rightarrow s^{6h}$\quad if $q_i:INC(x)\in\{Q\},\, x\neq 2$ \\
		& $(s^{8h})^{\ast}s^{16h+4(h+i)}/s^{12h+4i}\rightarrow s^{6h+4i}$\quad if $q_i:DEC(2)\in\{Q\}$ \\ 
		& $s^{8h+4(h+i)}/s^{4(h+i)}\rightarrow s^{4(h+i)}$\quad if $q_i:DEC(2)\in\{Q\}$ \\  
		& $(s^{8h})^{\ast}s^{4(h+i)}/s^{4(h+i)}\rightarrow s^{2h}$\quad if $q_i:DEC(x)\in\{Q\},\, x\neq 2$ \\ 
 \hline
$\sigma_4$	& $s^{8h+1}/s^{8h}\rightarrow s^{8h-1},$\qquad $s^{8h+2}/s^{8h}\rightarrow s^{8h-1},$ \qquad  $s^{8h+3}\rightarrow s^{2h}$  \\ 
		& $(s^{8h})^{\ast}s^{4(h+i)}/s^{4(h+i)}\rightarrow s^{4(h+i)}$\quad if $q_i:INC(3)\in\{Q\}$ \\ 
		& $(s^{8h})^{\ast}s^{4(h+i)}/s^{8h+4(h+i)}\rightarrow s^{6h}$\quad if $q_i:INC(x)\in\{Q\},\, x\neq 3$ \\
		& $(s^{8h})^{\ast}s^{16h+4(h+i)}/s^{12h+4i}\rightarrow s^{6h+4i}$\quad if $q_i:DEC(3)\in\{Q\}$ \\ 
		& $s^{8h+4(h+i)}/s^{4(h+i)}\rightarrow s^{4(h+i)}$\quad if $q_i:DEC(3)\in\{Q\}$ \\  
		& $(s^{8h})^{\ast}s^{4(h+i)}/s^{4(h+i)}\rightarrow s^{2h}$\quad if $q_i:DEC(x)\in\{Q\},\, x\neq 3$ \\  \hline
$\sigma_5$	& $s\rightarrow\lambda,$ \qquad   $s^{2h}\rightarrow \lambda,$\qquad   $s^{6h}\rightarrow \lambda,$\qquad   $s^{4(h+i)}\rightarrow \lambda,$\qquad    $s^{6h+4i}\rightarrow \lambda,$\qquad  $s^{2}\rightarrow s$  \\ \hline  
\end{tabular}
\end{center}
\caption{This table gives the rules for each of the neurons of $\Pi_{C_3}$.}\label{tab:neurons of Extended SNP 3-counter}
\end{table}

\begin{table}[h]
\begin{center}
\begin{tabular}{c@{\;\;\;}|@{\;\;\;}l}
neuron       	& rules  \\ \hline
$\sigma_1$   	& $s^{8h+1}/s^{8h}\rightarrow s^{8h},$\qquad   $s^{8h+2}/s^{8h-1}\rightarrow s^{4h+3},$ \qquad   $s^{8h+3}\rightarrow \lambda,$ \qquad $s^{2}\rightarrow \lambda,$\qquad   $s\rightarrow \lambda$  \\ 
		& $s^{16h+4i}\rightarrow s^{12h+4l},$\qquad  $s^{10h+4i}\rightarrow s^{4(h+l)},$\qquad if $l<h$ \\ 
		& $s^{16h+4i}\rightarrow s^{12h+5},$\qquad  $s^{10h+4i}\rightarrow s^{4h+5},$\qquad if $l=h$ \\
& $s^{8h+4i}\rightarrow s^{4(h+k)},$\qquad  if $k\neq h$ \\
& $s^{8h+4i}\rightarrow s^{4h+5},$\qquad  if $k=h$ \\\hline
$\sigma_2$ 	& $(s^{8h})^{\ast}s^{4(h+i)}/s^{4(h+i)}\rightarrow s^{4(h+i)}$\quad if $q_i:INC(1)\in\{Q\}$ \\ 
		& $(s^{8h})^{\ast}s^{4(h+i)}/s^{8h+4(h+i)}\rightarrow s^{12h}$\quad if $q_i:INC(2)\in\{Q\}$ \\
		& $(s^{8h})^{\ast}s^{16h+4(h+i)}/s^{12h+4i}\rightarrow s^{6h+4i}$\quad if $q_i:DEC(1)\in\{Q\}$ \\ 
		& $s^{8h+4(h+i)}/s^{4(h+i)}\rightarrow s^{4(h+i)}$\quad if $q_i:DEC(1)\in\{Q\}$ \\  
		& $(s^{8h})^{\ast}s^{4(h+i)}/s^{4(h+i)}\rightarrow s^{4h}$\quad if $q_i:DEC(2)\in\{Q\}$ \\ \hline
$\sigma_3$	& $s^{2}/s\rightarrow s,$\qquad   $s^{16h+1}/s^{8h}\rightarrow s^{8h},$ \qquad $(s^{8h})^{\ast}s^{20h+5}/s^{12h}\rightarrow s^{2}$  \\
		& $(s^{8h})^{\ast}s^{16h+5}/s^{8h}\rightarrow s,$\qquad    $s^{8h+5}\rightarrow s^2,$ \qquad   $s^{12h+5}\rightarrow s^2$ \\
		& $(s^{8h})^{\ast}s^{4(h+i)}/s^{4(h+i)}\rightarrow s^{4(h+i)}$\quad if $q_i:INC(2)\in\{Q\}$ \\ 
		& $(s^{8h})^{\ast}s^{4(h+i)}/s^{8h+4(h+i)}\rightarrow s^{12h}$\quad if $q_i:INC(1)\in\{Q\}$ \\
		& $(s^{8h})^{\ast}s^{16h+4(h+i)}/s^{12h+4i}\rightarrow s^{6h+4i}$\quad if $q_i:DEC(2)\in\{Q\}$ \\ 
		& $s^{8h+4(h+i)}/s^{4(h+i)}\rightarrow s^{4(h+i)}$\quad if $q_i:DEC(2)\in\{Q\}$ \\  
		& $(s^{8h})^{\ast}s^{4(h+i)}/s^{4(h+i)}\rightarrow s^{4h}$\quad if $q_i:DEC(1)\in\{Q\}$ \\  \hline
$\sigma_5$	&$s^{8h}\rightarrow \lambda,$\qquad $s^{12h}\rightarrow \lambda,$\qquad $s\rightarrow\lambda,$ \qquad   $s^{4h}\rightarrow \lambda,$\qquad   $s^{6h+4i}\rightarrow \lambda,$\qquad  $s^{4(h+i)}\rightarrow \lambda,$\qquad   $s^{2}\rightarrow s$  \\ \hline  
\end{tabular}
\end{center}
\caption{This table gives the rules for each of the neurons of $\Pi_{C_2}$.}\label{tab:neurons of Extended SNP 2-counter}
\end{table}

\begin{table}[h]
\begin{center}
\begin{tabular}{c@{\;\;\;}|@{\;\;\;}l}
neuron       	& rules  \\ \hline
$\sigma_1$				& $s\rightarrow s$,  \\ \hline
$\sigma_2$				& $s^{41}/s\rightarrow s$,\qquad $s^{43}/s^3\rightarrow s$,\qquad $s^{44}/s^{25}\rightarrow s$,\qquad $s^{22}/s^3\rightarrow s$,\qquad $s^{23}/s^{5}\rightarrow s$, \\ & $s^{21(h+i)-1}/s^{5}\rightarrow s$,\qquad $s^{21(h+i)-5}/s^{3}\rightarrow s$,\qquad  $s^{21(h+i)-7}/s^{2}\rightarrow s,$\qquad $s^{21(h+i)-8}/s^{3}\rightarrow s$\\
& $s^{21(h+i)-10}/s^{5}\rightarrow s$,\qquad $s^{42h+1}/s\rightarrow s,$\qquad $s^{\ast}s^{42h+3}/s\rightarrow s$\\
& $s^{21(h+i)+1}/s^{4}\rightarrow s$\quad if $q_i:INC \in\{Q\}$\\
& $s^{21(h+i)-2}/s^{21(h+i-l)+1}\rightarrow s$\quad if $q_i:INC(1) \in\{Q\}$\\
& $s^{21(h+i)-2}/s^{21(h+i-l)+2}\rightarrow s$\quad if $q_i:INC(x) \in\{Q\},\;x\neq 1$\\
& $s^{21(h+i)-2}/s^{21(h+i-k)-1}\rightarrow s$\quad if $q_i:DEC \in\{Q\}$\\
& $s^{21(h+i)+1}/s^{5}\rightarrow s$\quad if $q_i:DEC(1)\in\{Q\}$\\
& $s^{21(h+i)+1}/s^{6}\rightarrow s$\quad if $q_i:DEC(x)\in\{Q\},\;x\neq 1$\\ 
& $s^{21(h+i)-11}/s^{6}\rightarrow s$\quad if $q_i:DEC(1)\in\{Q\}$\\
& $s^{21(h+i)-11}/s^{5}\rightarrow s$\quad if $q_i:DEC(x)\in\{Q\},\;x\neq 1$\\
& $s^{21(h+i)-13}/s^{21(h+i-l)-6}\rightarrow s$\quad if $q_i:DEC(1)\in\{Q\}$\\
& $s^{21(h+i)-13}/s^{21(h+i-l)-7}\rightarrow s$\quad if $q_i:DEC(x)\in\{Q\},\;x\neq 1$\\ 
\hline
$\sigma_3$				& $s^{41}/s\rightarrow s$,\qquad $s^{43}/s^3\rightarrow s$,\qquad $s^{44}/s^{31}\rightarrow s$,\qquad $s^{16}/s^{3}\rightarrow s$, \qquad $s^{21(h+i)-1}/s^{5}\rightarrow s$\\
& $s^{21(h+i)+5}/s^{11}\rightarrow s$,\qquad $s^{21(h+i)-5}/s^{3}\rightarrow s$, \qquad $s^{21(h+i)-7}/s^{2}\rightarrow s$,\qquad $s^{21(h+i)-8}/s^{3}\rightarrow s$ \\ 
& $s^{21(h+i)-11}/s^{6}\rightarrow s$\ \qquad  $s^{21(h+i)-13}/s^{21(h+i-l)-6}\rightarrow s$,\qquad $s^{21(h+i)+4}/s^{21(h+i-k)+5}\rightarrow s$\\
&$s^{42h+1}/s\rightarrow s$,\qquad $s^{\ast}s^{42h+3}/s\rightarrow s$\\
& $s^{21(h+i)+1}/s^{21(h+i-l)+6}\rightarrow s$\quad if $q_i:INC(1)\in\{Q\}$ \\ 
& $s^{21(h+i)+1}/s^{4}\rightarrow s$\quad if $q_i:INC(x)\in\{Q\},\;x\neq 1$\\ 
& $s^{21(h+i)-2}/s^{21(h+i-l)+2}\rightarrow s$\quad if $q_i:INC(x) \in\{Q\}$\\ 
& $s^{21(h+i)-2}/s^{21(h+i-k)-1}\rightarrow s$\quad if $q_i:DEC \in\{Q\}$\\
& $s^{21(h+i)+1}/s^{6}\rightarrow s$\quad if $q_i:DEC(x)\in\{Q\},\;x\neq 1$\\
& $s^{21(h+i)-10}/s^{5}\rightarrow s$\quad if $q_i:DEC(1)\in\{Q\}$\\
& $s^{21(h+i)-10}/s^{21(h+i-l)+5}\rightarrow s$\quad if $q_i:DEC(x)\in\{Q\},\;x\neq 1$\\
\hline
\end{tabular}
\end{center}
\caption{This table gives the rules for neurons $\sigma_1$ to $\sigma_3$ of $\Pi'_{C_3}$.}\label{tab:neurons of SNP I}
\end{table}

\begin{table}[h]
\begin{center}
\begin{tabular}{c@{\;\;\;}|@{\;\;\;}l}
neuron       	& rules  \\ \hline
$\sigma_4$				& $s^{41}/s\rightarrow s$,\qquad $s^{43}/s^3\rightarrow s$,\qquad $s^{44}/s^{31}\rightarrow s$,\qquad $s^{16}/s^{3}\rightarrow s$,\qquad $s^{21(h+i)-1}/s^{5}\rightarrow s$  \\ 
& $s^{21(h+i)+5}/s^{11}\rightarrow s$,\qquad $s^{21(h+i)-5}/s^{3}\rightarrow s$,\qquad $s^{21(h+i)-8}/s^{3}\rightarrow s$\\ 
& $s^{21(h+i)-10}/s^{21(h+i-l)+5}\rightarrow s,$ \qquad$s^{21(h+i)+4}/s^{21(h+i-k)+5}\rightarrow s$,\qquad $s^{42h+1}/s\rightarrow s$\\
& $s^{21(h+i)+1}/s^{21(h+i-l)+6}\rightarrow s$\quad if $q_i:INC(x)\in\{Q\},\;x\neq 3$\\ 
& $s^{21(h+i)+1}/s^{4}\rightarrow s$\quad if $q_i:INC(3)\in\{Q\}$\\ 
& $s^{21(h+i)-2}/s^{21(h+i-l)+2}\rightarrow s$\quad if $q_i:INC(x) \in\{Q\}$\\ 
& $s^{21(h+i)-2}/s^{21(h+i-k)-1}\rightarrow s$\quad if $q_i:DEC \in\{Q\}$\\
& $s^{21(h+i)+1}/s^{6}\rightarrow s$\quad if $q_i:DEC(3)\in\{Q\}$\\
& $s^{21(h+i)-7}/s^{21(h+i-l)+10}\rightarrow s$\quad if $q_i:DEC(3)\in\{Q\}$\\
& $s^{21(h+i)-7}/s^{2}\rightarrow s$\quad if $q_i:DEC(x)\in\{Q\},\;x\neq 3$\\
\hline
$\sigma_5$				& $s^{41}/s\rightarrow s$,\qquad $s^{43}/s^3\rightarrow s$,\qquad $s^{44}/s^{31}\rightarrow s$,\qquad $s^{16}/s^{3}\rightarrow s$\\
&$s^{21(h+i)+5}/s^{11}\rightarrow s$,\qquad $s^{21(h+i)-5}/s^{3}\rightarrow s$,\qquad  $s^{21(h+i)-7}/s^{21(h+i-l)+10}\rightarrow s$ \\ 
&$s^{21(h+i)+4}/s^{21(h+i-k)+5}\rightarrow s$,\qquad $s^{42h+1}/s\rightarrow s$\\
& $s^{21(h+i)+1}/s^{21(h+i-l)+6}\rightarrow s$\quad if $q_i:INC\in\{Q\}$ \\ 
\hline
$\sigma_6,\sigma_7$			& $s^{41}/s\rightarrow s$,\qquad $s^{43}/s^3\rightarrow s$,\qquad $s^{44}/s^{31}\rightarrow s$,\qquad $s^{16}/s^{3}\rightarrow s$,  \\
&$s^{21(h+i)+5}/s^{11}\rightarrow s$,\qquad $s^{21(h+i)-5}/s^{3}\rightarrow s$,\qquad  $s^{21(h+i)-7}/s^{21(h+i-l)+10}\rightarrow s$ \\ 
&$s^{21(h+i)+4}/s^{21(h+i-k)+5}\rightarrow s$\\
& $s^{21(h+i)+1}/s^{21(h+i-l)+6}\rightarrow s$\quad if $q_i:INC\in\{Q\}$ \\ \hline
$\sigma_{8}$				& $(s^{6})^\ast s^{11}/s^{6}\rightarrow s$,\qquad $(s^{6})^\ast s^{13}/s\rightarrow s$,\qquad $s^{7}\rightarrow \lambda$, \\ 
& $(s^{6})^\ast s^{8}/s^{8}\rightarrow s$,\qquad $(s^{6})^\ast s^{9}/s^{9}\rightarrow s$,\\ 
\hline
$\sigma_{9}$				& $(s^{6})^\ast s^{10}/s^{6}\rightarrow s$,\qquad $(s^{6})^\ast s^{7}/s^{7}\rightarrow s$,\qquad  $(s^{6})^\ast s^{14}/s^{2}\rightarrow s$, \\ 
& $s^{8}\rightarrow \lambda$,\qquad $(s^{6})^\ast s^{9}/s^{9}\rightarrow s$, \\ 
\hline
$\sigma_{10}$				& $(s^{6})^\ast s^{10}/s^{6}\rightarrow s$,\qquad $(s^{6})^\ast s^{11}/s^{6}\rightarrow s$,\qquad $(s^{6})^\ast s^{7}/s^{7}\rightarrow s$,\\  & $(s^{6})^\ast s^{8}/s^{8}\rightarrow s$,\qquad $(s^{6})^\ast s^{15}/s^{3}\rightarrow s$,\qquad $s^{9}\rightarrow \lambda$,\qquad $s^{2}\rightarrow \lambda$ \\ 
\hline
$\sigma_{11}$				& $s^7\rightarrow \lambda$,\qquad  $s^6\rightarrow \lambda$,\qquad  $s\rightarrow \lambda$,\qquad $s^{11}/s\rightarrow s$,\qquad $(s^3)^{\ast}s^{14}/s\rightarrow s$,\\
& $s^{4}\rightarrow \lambda$,\qquad $s^{2}\rightarrow \lambda$,\qquad $s^{3}\rightarrow \lambda$,\qquad $s^{10}\rightarrow s$\\ 
\hline
$\sigma_{12},\sigma_{15}$				& $(s^{3})^\ast s^{4}/s^{3}\rightarrow s$,\qquad $s\rightarrow s$,  \\ 
\hline
$\sigma_{13},\sigma_{14},\sigma_{16},\sigma_{17}$		& $(s^{3})^\ast s^{4}/s^{3}\rightarrow s$,\qquad $s\rightarrow \lambda$,  \\
\hline
\end{tabular}
\end{center}
\caption{This table gives the rules for neurons $\sigma_4$ to $\sigma_{17}$ of $\Pi'_{C_3}$.}\label{tab:neurons of SNP II}
\end{table}

\end{document}